\documentclass[conference]{IEEEtran}
\IEEEoverridecommandlockouts
\usepackage{cite}
\usepackage{amsmath,amssymb,amsfonts}
\usepackage{algorithm}
\usepackage{algorithmic}
\usepackage{graphicx}
\usepackage{textcomp}
\usepackage{xcolor}
\usepackage{listings}
\usepackage{hyperref}
\usepackage{float}
\usepackage{stfloats}
\usepackage{placeins}

\def\BibTeX{{\rm B\kern-.05em{\sc i\kern-.025em b}\kern-.08em
    T\kern-.1667em\lower.7ex\hbox{E}\kern-.125emX}}

\begin{document}

\newcommand{\dslhype}{\textsc{DSLHyPE}}
\newcommand{\exahype}{ExaHyPE}

\title{\dslhype---a DSL kernel language for the Exascale Hyperbolic PDE Engine \exahype
  \thanks{
This work, and notably Maurice's and Timothy's work, have received funding through the eCSE project ARCHER2-eCSE11-2 \emph{ExaHyPE-DSL}.
Tobias appreciates the support by Intel through their Intel Academic Centre of Excellence supporting Timothy's PhD studentship. 
The performance studies and analyses, i.e.~Thomas' contribution, have been enabled via funding through the UKRI Digital Research Infrastructure Programme under grant UKRI1801 (SHAREing) and under grant number UKRI/ST/B000293/1 (HAI-End).
This work has made use of the HPC Hardware Lab @ Durham. 
  }
}

\author{
  \IEEEauthorblockN{1\textsuperscript{st} Timothy J.R. Stokes}
  \IEEEauthorblockA{\textit{Department of Computer Science} \\
  \textit{Durham University}\\
    Durham, United Kingdom \\
    0009-0008-8607-6343
  }
  \and
  \IEEEauthorblockN{2\textsuperscript{nd} Nick Brown}
  \IEEEauthorblockA{\textit{EPCC} \\
  \textit{University of Edinburgh}\\
    Edinburgh, United Kingdom \\
    0000-0003-2925-7275}
  \and
  \IEEEauthorblockN{3\textsuperscript{rd} Thomas A. Flynn}
  \IEEEauthorblockA{\textit{Advanced Research Computing} \\
  \textit{Durham University}\\
    Durham, United Kingdom \\
    0000-0001-9800-594X
  }
  \and
  \IEEEauthorblockN{4\textsuperscript{th} Maurice Jamieson}
  \IEEEauthorblockA{\textit{EPCC} \\
  \textit{University of Edinburgh}\\
    Edinburgh, United Kingdom \\
    0000-0003-1626-4871
  }
  \and
  \IEEEauthorblockN{5\textsuperscript{th} Tobias Weinzierl}
  \IEEEauthorblockA{\textit{Department of Computer Science} \\
  \textit{Institute for Data Science}\\
    \textit{Durham University}\\
    Durham, United Kingdom \\
    0000-0002-6208-1841
  }
}

\maketitle

\begin{abstract}
We introduce a bilingual domain-specific language (DSL) for modelling compute kernels within a generic solver for hyperbolic partial differential equations (PDEs). 
Users express PDE terms, i.e.~the underlying physics, in a familiar native language such as C or C++, while the numerical scheme is specified in a Python-embedded DSL, \dslhype. 
\dslhype's compiler lowers the Python description to MLIR and introduces a translation pass that integrates it with native code likewise mapped to MLIR. 
Our approach keeps the numerical representation and the physics implementation separate for as long as possible, while delegating optimization to the compiler through existing MLIR optimization passes.
This separation of concerns benefits researchers developing numerical schemes on top of existing PDE implementations or with applications involving nonlinear systems whose PDE terms must solve PDEs themselves.
%, and communities that prefer scientific programming languages other than Python. 
We demonstrate the feasibility of the approach using a gravitational-wave solver and a matter-evolution solver on x86 processors and H200 GPUs.

\end{abstract}


% \marginpar{8 pages w/o bib}
% \marginpar{Deadline 14 August 2026}

\section{Introduction}

%
% General context: kernels are expensive and need tailoring
%
\exahype~\cite{Reinarz:2020:ExaHyPE} is a general-purpose engine for simulating a wide range of wave phenomena, i.e.~hyperbolic partial differential equations (PDEs), that are discretized via explicit time-stepping methods.
In line with many hyperbolic PDEs simulation packages including~\cite{Deppe:2026:Spectre,SeisSol,GRChombo}, the \exahype~favours high-order discretizations, but also offers low-order Finite Volumes (FV) for phenomena that are prone to shocks.
The engine supports PDEs of the form

\begin{equation}
 \partial _t Q + \nabla F(Q) + B(Q) \cdot \nabla Q = S(Q)
 \label{equation:introduction:PDE}
\end{equation}

\noindent
in a problem-agnostic way via generic numerical compute kernels that are independent of $F(Q)$, $B(Q)$ and $S(Q)$,
even though advanced PDEs quickly require users to develop their own tailored numerics.
\exahype~s numerical astrophysics application \cite{Zhang:2025:ExaGRyPE} is one domain which relies upon custom Riemann solvers incorporating physical admissibility checks or differencing schemes with appropriate Kreiss-Oliger dissipation terms \cite{Kreiss:1973:Dissipation}.
These compute kernels must run on both CPUs and GPUs.
\exahype~hence needs a flexible, powerful approach to bring the actual physics, the PDE operators, and bespoke numerical schemes together. 
To manage complexity it is crucial that this approach is able to hides the low-level details of optimization and hardware porting.

%
% DSLs
% - embedded vs stand-alone
% - mathematical front-end plus translation rules vs.~separate
%
Domain-specific languages (DSLs) have a mainstream approach for providing such an interface.
Once tailored to the domain, in our case hyperbolic solvers, DSLs can provide appropriate levels of abstraction for domain and numerical scientists to work within, whilst delivering significant information around the intention of compute to the compiler itself upon which it can optimize and tune the resulting code.
They offer separation of concerns, efficiency plus developer productivity.
DSLs embedded within existing languages, such as Devito in Python for Seismology \cite{lange2016devito} or PSyclone in Fortran for weather and climate\cite{siso2023transforming} provide popular success stories. 
%Ultimately the value of a DSL is in the abstraction level which is provided to the end user for expressing their mathematical problems.

%further distinguished by whether they offer their own optimisation steps or sit on top of community translation tools such as Multi-Level IR (MLIR) \cite{Lattner:2021:MLIR}.
%Another way to characterise DSLs is the way they support expert knowledge of the disciplines involved, i.e.~domain sciences, mathematics and HPC computer science:
%Do they embed disciplinary knowledge within their translation stack, e.g.~do they offer a strict mathematical front-end and apply mathematical transformations under the hood, but cover the whole abstraction stack from mathematical formulation down to efficient multi-platform code, or do they focus on one particular aspect such as the efficient treatment of already discretised and linearised PDE operators?
%Finally, DSLs can be characterised by their scope and integration into the host domain (code).
%Some DSLs hide whole frameworks including MPI data exchange, load balancing and offloading decisions behind the language \cite{Bauer:2019:PyStencils}, while others focus ``only'' on compute kernel or glue code generation. 

Modern compiler technologies, such as MLIR \cite{Lattner:2021:MLIR}, have opened up the ability for developers to quickly develop their tailored DSL,
while leveraging an existing, rich compiler ecosystem.  In this paper, we introduce a \dslhype~for \exahype.
\dslhype{} enables programmers to express their numerics in an abstract manner in Python, with the physical PDE terms still programmed in C, C++ (\exahype's core language) or SymPy. 
\dslhype's compiler takes the numerical recipe of how to update a cell, combines it with the user-provided physics functions,
and then lowers the merger into the Multi-Level Intermediate Representation (MLIR) \cite{Lattner:2021:MLIR} and eventually 
LLVM-IR which is passed on into the LLVM compiler stack \cite{Lattner:2004:LLVM}.

The novel contributions of this paper comprise
\begin{itemize}
  \item the concept of a bilingual DSL which integrates a codebase written in multiple languages (e.g.~Python and C);
  \item highlighting MLIR transformation passes to deliver performance on the CPU and GPU, leveraging a range of LLVM sub-projects;
  \item identifying key optimisations missing within the MLIR ecosystem to deliver performance on the CPU and GPU.
\end{itemize}

\noindent
The remainder of this paper is organized as follows:
we first introduce our DSL architecture and its language ingredients in Section~\ref{section:challenge}.
From hereon, Section~\ref{section:pipeline} discusses our translation pipeline which puts special emphasis on the integration of user-provided code and the realization with the MLIR/LLVM ecosystem.
Benchmarking this arrangement (Sections~\ref{section:benchmarks}) enables us to explore performance of the current \dslhype and then sketch bespoke code optimization passes in Section~\ref{section:optimisation} before highlighting related work in Section~\ref{section:related-work}. 
Section~\ref{section:conclusion} concludes this work and highlights future work.

\section{DSLHyPE: A bilingual DSL for \exahype}
\label{section:challenge}

Compute kernels in \exahype~are formulated over a mesh of hypercubes. This means, for higher-order polynomials, each cube hosts Gauss-Lagrangian or Gauss-Lobatto shape functions, while Finite Volume (FV) and Finite Difference (FD) schemes embed regular $p \times p \ (d=2)$ or $p \times p \times p \ (d=3)$ Cartesian meshes (patches) into the cells.
\exahype's core compute kernel takes a cubic or square cell within a computational mesh and yields its solution at the new time step:

\begin{equation}
 \mathcal{K}_{\Delta t}: \mathbb{R}^{\hat p^d(N+M)} \mapsto \mathbb{R}^{p^dN}.
 \label{equation:challenge:cell-kernel}
\end{equation}

\noindent
The input/output values represent the quantities of $Q \in \mathbb{R}^{N+M}$ in \eqref{equation:introduction:PDE}. 
$N$ quantities encode information such as pressure, velocity or density, while the $M$ material parameters are fixed.
They do not evolve in time.

For FV and FD schemes, each cell's embedded mesh is augmented by a halo layer of width $h$ ($\hat p=p+2h$).
Runge-Kutta schemes require multiple cell update calls interwoven with different time step size $\Delta t$ choices according to the Butcher tableau, as well as synchronization steps with face-adjacent cells.
In all cases, \exahype{} typically does not invoke \eqref{equation:challenge:cell-kernel} per grid cell, but instead accepts $K\geq 1$ cells and advances all of them concurrently by $\Delta t$ \cite{Li:2022:DynamicTaskFusion}
through a kernel $\tilde{ \mathcal{K} }_{\Delta t, K}$.

\subsection{Kernel composition}

The per-cell operator \eqref{equation:challenge:cell-kernel} can be written down as a sequence of $S$ sub-operators

\begin{equation}
 \mathcal{K}_{\Delta t}(x) = \left( \mathcal{K}_{\Delta t}^{(S)} \circ \ldots \circ \mathcal{K}_{\Delta t}^{(2)} \circ \mathcal{K}_{\Delta t}^{(1)} \right) (x),
 \label{equation:challenge:operator-composition}
\end{equation}

\noindent
where some sub-operators consume the outcomes of previous steps, some react solely to (parts of) $x$ encoding all $Q$ values over the cell, and some combine a mixture of both.

For Finite Volumes, a first operator $\mathcal{K}_{\Delta t}^{(1)}$ might evaluate a source term $S(Q)$ from \eqref{equation:introduction:PDE} over every degree of freedom (patch voxel) within the cell.
The next operator $\mathcal{K}_{\Delta t}^{(2)}$ takes the outcome of this and adds it to the input, subject to a scaling with the time step size $\Delta t$.
After that, an operator $\mathcal{K}_{\Delta t}^{(3)}$ evaluates the fluxes $F$ in \eqref{equation:introduction:PDE} over all horizontal faces, for which it first reconstructs a $Q$ from the upper and lower voxel using a linear combination, i.e.~a stencil.
Eventually, all results feed into an updated cell state.

Nonlinearities in the arising scheme can result either from a nonlinear combination of input data in \eqref{equation:challenge:operator-composition},
or from the evaluation of the terms in \eqref{equation:introduction:PDE}, which internally might run nonlinear calculations over ``their'' $Q$ input.

\subsection{Data layout}

$x$ holds the sequence of $Q$ values over a cell.
They are arranged as an array of structs (AoS) that is stored contiguously in memory.
The overall compute kernel $\mathcal{K}_{\Delta t}$ yields an output of $N$ quantities which are stored as AoS, too.

Most PDE operators are plain functions that take an input of $N+M$ weights, each of which are double precision values. Consequently, a straightforward implementation of the terms from \eqref{equation:introduction:PDE} operates over contiguous fragments of the AoS input.
However, \dslhype~does not constrain what input functions expect.
Flux realizations for example might also require directional derivatives, while functions evaluating the maximum eigenvalue return only scalars.
%Independent of the variable count, all input and output of functions are logically AoS, although a compiler might internally want to rearrange data layouts to obtain better performance \cite{Radtke:2025:SoAtoAoS}.

The data per cell are contiguous, but the operator $\tilde{ \mathcal{K} }_{\Delta t, K}$ does not operate on a contiguous set of unknowns for an arbitrary set of patches.
It is given $K$ chunks of data, where each chunk is contiguous but the sequence is not.
We may interpret this as an array of AoS (AoAoS).

\subsection{Kernel syntax}
\label{sec:dsl_design}
\begin{algorithm}[htb]
  \begin{center}  
    \footnotesize
    \begin{lstlisting}[breaklines=true,numbers=left]
def computeKernel(cellData: "CellDescription") -> "void":
  """!
  The values N, M are defined in the Python 
  context. This is a Finite Volume kernel with a 
  halo of h around each cell's grid. 
  """
  K = Integer("cellData.numberOfCells")
  ...

  input = DataBlock([[0,N+M], [-h,p+h], [-h,p+h], [0,K]], "patchData.QIn")
  output = DataBlock([[0,N+M], [0,p], [0,p], [0,K]], "patchData.QIn")
  
  # Copy data over
  output = input[:, 0:p, 0:p]
  
  # Flux evaluation
  flux_x = DataBlock([[0,N],[-h,p+h],[0,p],[0,N]], eval("F", input[:,-h:p+h,0:p], ...))
  
  # Add outcome to ouptut 
  output[0:N] = output[0:N] + 0.5 * p * dt / dx * (flux_x[:,-1:p-1,0:p] - flux_x[:,1:p+1,0:p])
  ...
    \end{lstlisting}
  \end{center}
%sources = DataBlock([[0,N], [0,p], [0,p], [0,K]], eval(S(input[:, 0:p, 0:p], x, ...)))
%output = input[:, 0:p, 0:p] + dt * sources
  \caption{
    Fragment 2d compute kernel in \dslhype.
    \label{algorithm:compute-kernel}
  }  
\end{algorithm}

\begin{figure}[htb]
  \begin{center}
    \includegraphics[width=0.3\textwidth]{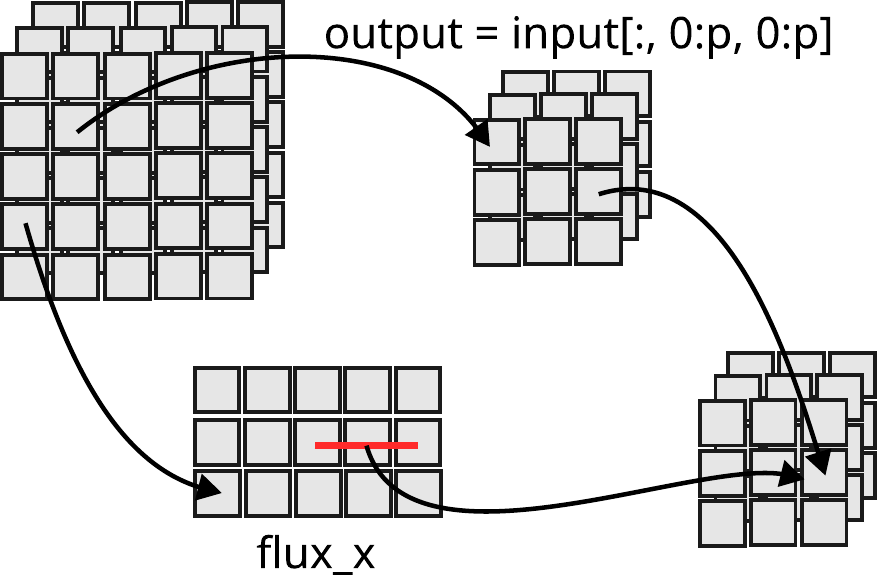}
  \end{center}
  \caption{
    Schematic illustration of the compute kernel from Algorithm~\ref{algorithm:compute-kernel},
    where we assume that the kernel is initialized with $h=1$ and $p=3$, while $K=3$.
    \label{figure:compute-kernel}
  }  
\end{figure}

While we assume that a user of the engine provides PDE terms from \eqref{equation:introduction:PDE} as functions written in C, C++ or through SymPy's code generation,
our work proposes that the compute kernels are written in a Python-esque stencil formulation (Algorithm~\ref{algorithm:compute-kernel}),
which is syntactically valid Python and hence can be represented by Python's concrete syntax tree (CST).

The compute kernel, $\tilde{ \mathcal{K} }_{\Delta t, K}$, is defined over attributes that exist in the underlying \exahype~C++ code, \texttt{patchData} in Algorithm \ref{figure:compute-kernel}, which aggregates further data. 
Encoding its type as a string instructs the Python DSL that the kernel will be mapped onto a (compiled) C function over a data type \texttt{CellDescription}. The benefit of making all attributes explicitly visible within the \dslhype~function means that the DSL can also be used with non-\exahype~codes.

The example kernel in Algorithm \ref{figure:compute-kernel} introduces a variable \texttt{K}  at line 7, which is the cell count. This variable links to the underlying C++ context and is read from the input arguments of the defined kernel function.
Other variables without an explicit binding (e.g.~\texttt{N}, \texttt{M}, \texttt{p}, \texttt{h} in Algorithm \ref{figure:compute-kernel}) are assumed to be global and static.

The DSL represents all calculations as mappings between \texttt{DataBlock}s, which are arrays with a range, a dimension and size, provided by the programmer. 
There are three flavors of data blocks.
Firstly those tied to existing arrays for example the input struct \texttt{cellData} in Algorithm \ref{figure:compute-kernel} holds two arrays \texttt{QIn} and \texttt{QOut}, which we create Python \texttt{DataBlock} aliases for.
Secondly, datablocks can be constructed through function evaluations, for example \texttt{flux\_x} at line 17 in Algorithm \ref{figure:compute-kernel} is a data block where each entry is determined by a function evaluation of an external function \texttt{F}. 
Thirdly. a data block can be constructed by a calculation which is expressed via stencil operations or the application of a global matrix to the input block (not shown).

Data blocks that are not explicitly tied to input data are effectively temporary variables within the kernel and while their size must be explicitly provided, this does not have to be a compile-time constant. An example of this is the variable \texttt{K} in Algorithm \ref{figure:compute-kernel}, holding the cell count $K$, which is a runtime parameter.
Our DSL abstraction also provides flexible index ranges where indices do not need to start from 0. This can aid in programmability and can be seen in Algorithm \ref{figure:compute-kernel} where \texttt{QIn} is (logically) an array of size $(N+M) \cdot (p+2h) \cdot (p+2h) \cdot K$, although the middle two indices run from $-h$ to $p+h-1$.

Once \texttt{input} and \texttt{output} are tied to existing arrays at lines 10 and 11 of Algorithm \ref{figure:compute-kernel}, the next step copies the input data, minus its halo, into \texttt{output} at line 14. Next a new data range evaluates a user-provided function over the input array at line 17, and the outcome of this calculation is added to \texttt{output} at line 20, providing a simple linear combination of values logically left and right within the ranges. 
Left and right accesses are realized via index shifts.
% for non-zero indexed arrays.

Algorithm \ref{figure:compute-kernel} mirrors Equation \eqref{equation:challenge:operator-composition}, constructing the output as a sequence of compute steps over temporary data blocks as is illustrated in Figure~\ref{figure:compute-kernel}. 
Concurrency between these steps is not expressed in the kernel syntax, but each step corresponds to an iteration over the data range and each range element is only assigned once. This single-assignment \cite{Scholz:1996:SAC} mirrors functional programming and makes data dependencies between calculation steps plain and explicit. %The notation over data ranges looping over range elements, unless explicitly fixed, bears similarities with Einstein's sum convention.

\section{\dslhype's MLIR compilation pipeline}
\label{section:pipeline}

The \dslhype~approach is to leverage MLIR \cite{Lattner:2021:MLIR} where possible.
MLIR provides a series of Intermediate Representation (IR) dialects, at different levels of abstraction, and transformations between these. Ultimately, it generates LLVM-IR which is then compatible with LLVM \cite{Lattner:2004:LLVM} backends, such as those targeting CPUs or GPUs.
A major benefit of MLIR is that it is a framework where developers can add their own dialects and transformations, leveraging the wealth of existing infrastructure that already exists in the lowering to executable code. 
For \dslhype, we leverage a combination of our own and existing transformation passes to transform programmer's code in our DSL into CPU or GPU binaries.

\subsection{Lowering into a sequence-of-nested-loops representation}

\begin{figure*}[thb]
  \begin{center}
    \includegraphics[width=\linewidth]{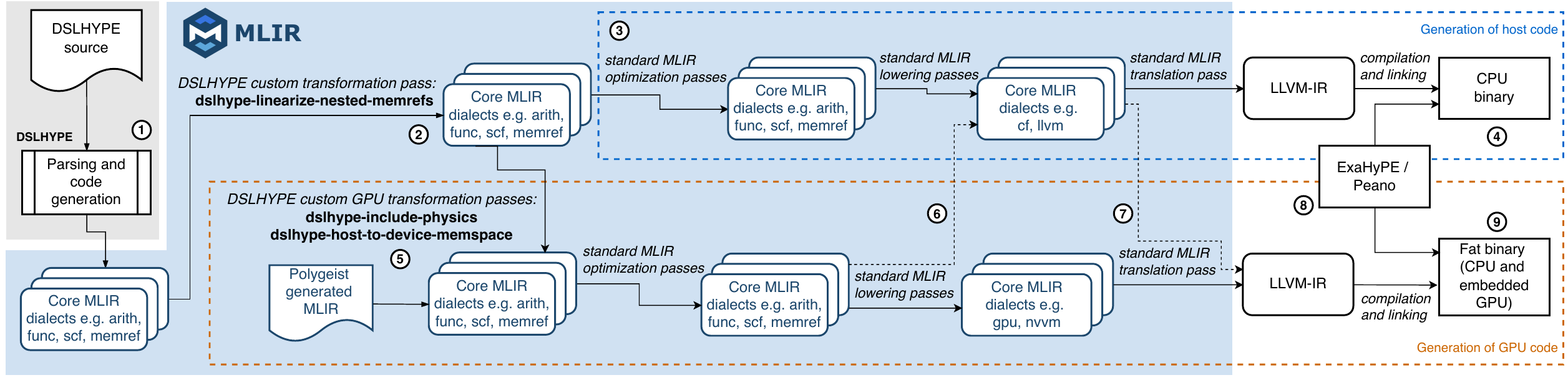}
  \end{center}
  \caption{
    \dslhype~MLIR code generation approach for CPU and GPU, including standard dialects and custom pipeline passes.
    \label{figure:mlir-piplines}
  }
\end{figure*}

As \dslhype~uses Python as a frontend, the initial parsing of the source into tokens is carried out via Python's \emph{ast} module \cite{Webpage:2026:AST} which produces a concrete syntax tree (CST). 
There are then two potential routes that can be followed, the first is a pretty printer that operates over this tree to run depth-first through the parsed data structure. The printer propagates array ranges top-down in-order along the source code encoded by the CST, and maps the patch assignments involving index notations onto plain loops, while each step with a data block of the algorithm introduces a new (temporary) array. 
Whilst this simple approach is optimized, it acts as a useful baseline.

Our main compiler pipeline lowers into MLIR where the CST is walked and upstream MLIR dialects, such as \texttt{arith} for arithmetic operations, \texttt{scf} for structured control flow and \texttt{memref} for memory allocation, are generated. At this stage the IR has a similar structure to the pretty-printed C++ code, a series of nested loops, but crucially we are then able to undertake a range of optimizations. As illustrated in Figure \ref{figure:mlir-piplines}, a mixture of existing MLIR transformations and our own bespoke passes are then leveraged as the IR is progressively lowered to target either the CPU ({\large \textcircled{\small 4}} in Figure \ref{figure:mlir-piplines}) or GPU ({\large \textcircled{\small 9}} in Figure \ref{figure:mlir-piplines}) by using different passes in their respective lowering pipelines ({\large \textcircled{\small 3}} and {\large \textcircled{\small 5}}). 

\subsection{Integrating with user-defined routines}

At the core of \dslhype's compilation approach is the ability for the description of the physics and the numerical scheme to sit together as equal class citizens.
The major challenge is how to integrate these because the description of the PDE is typically in C/C++ whereas the numerics are given by the Python CST. 
This mixture makes aggressive cross-function optimization and vectorisation challenging as long as the user functions remain in separate compilation units,
while it becomes notably a hurdle GPUs since MLIR's GPU passes require the IR to be present.
It is not possible to simply call out to kernels compiled elsewhere. 
Therefore, the user code has to be inlined into MLIR automatically and early throughout the translation:
\dslhype's compiler must be able to unify two languages and views of the user's code into a single IR.

Considering this PDE code can be large and complex, it is not appropriate to re-write machinery to parse it.
Instead, the \emph{dslhype-include-physics} pass ({\large \textcircled{\small 5}} in Figure \ref{figure:mlir-piplines}) firstly uses Polygeist \cite{Moses:2021:Polygeist,Webpage:2026:Polygeist} to transform the user's C++ PDE functions into IR. 
Secondly, it fuses both the compute kernels and user functions IR together to provide one unified representation. 
%In addition to supporting the targeting of GPUs, this fusion of compute and user functions also enables subsequent passes in the MLIR pipeline to optimize across the entire IR, rather than just focusing on one component.

\subsection{Memory flattening}
\label{subsection:pipeline:memory-flattening}

GPU offloading as well as some CPU optimizations rely heavily on appropriate memory arrangements. \exahype~typically works with a scattered AoAoS, but there is not a built-in guarantee in our DSL language that this is always the case. Consequently, all memory accesses have to be represented as nested, indirect memory accesses, i.e.~\texttt{memref<?xmemref<?xf64>>}, effectively \texttt{double**}, for AoAoS in the generated IR.

Indirect memory accesses are undesired due to their unsuitability for many optimizations including aggressive vectorization. 
We therefore introduce explicit gathering and scattering based around our \emph{dslhype-linearize-nested-memrefs} pass ({\large \textcircled{\small 2}} in Figure \ref{figure:mlir-piplines}). 
This pass maps all input and temporary memory onto one linear data space,  flattens nested memrefs to \texttt{memref<?x?xf64>}, and introduces the appropriate memory allocation and deallocation operations.

Formally, the kernel $\tilde{ \mathcal{K} }_{\Delta t}^{(S+1)}$ hosting $\mathcal{K}_{\Delta t}$ from \eqref{equation:challenge:operator-composition} is embedded into a prologue and epilogue

\begin{equation}
 \tilde{ \mathcal{K} }_{\Delta t}(x_1,..,x_K) = \left( \tilde{ \mathcal{K} }_{\Delta t}^{(S+1)} \circ \ldots \circ \tilde{ \mathcal{K} }_{\Delta t}^{(0)} \right) (x_1,..,x_K),
 \label{equation:challenge:gather-scatter}
\end{equation}

\noindent
where the first kernel $\tilde{ \mathcal{K} }_{\Delta t}^{(0)}$ takes the input vector $(x_1,x_2,\ldots,x_K)$ and maps it onto one continuous data space.
The last kernel $\mathcal{K}_{\Delta t}^{(S+1)}$ is the counterpart for the kernel outcome.

The \emph{dslhype-linearize-nested-memrefs} pass is not mandatory in producing correctly working code for the CPU, and it does introduce an overhead represented through $\mathcal{K}_{\Delta t}^{(0)}$ and $\mathcal{K}_{\Delta t}^{(S+1)}$, which subsequent translation passes have to compensate for.

\subsection{GPU offloading}

\dslhype~must produce efficient code which can run on both the CPU and GPU. 
While many modern accelerated systems offer virtual shared memory, HPC kernels often favor explicitly mapped memory for performance reasons.
This requires explicit data movement operations. 
To this end, \dslhype's \emph{dslhype-host-to-device-memspace} pass {\large \textcircled{\small 5}} updates all pointer- (\texttt{!llvm.ptr}) and memrefs-related operations to use temporary memory.
Furthermore, the pass augments $\mathcal{K}_{\Delta t}^{(0)}$ and $\mathcal{K}_{\Delta t}^{(S+1)}$ to explicitly allocate such temporary device (GPU) memory space and to map (copy) data to or from the accelerator, respectively. 
This pass inherently relies on a preceding \emph{dslhype-linearize-nested-memrefs}.

The pass makes the IR compatible with the upstream MLIR standard GPU lowering passes \cite{Webpage:2026:GPU} and GPU dialects, enabling the use of existing MLIR infrastructure to transform the IR into a GPU binary {\large \textcircled{\small 9}}.
At this point, the $S$ stages of the overall kernel are mapped onto $S$ kernel calls with each kernel updating $K$ cells in parallel, providing a chain of kernel executions on the GPU.

The MLIR GPU passes continue to generate host code which runs on the CPU, too (Figure~\ref{figure:mlir-piplines}, {\large \textcircled{\small 6}}). 
The dual-target capability is included in the IR that is lowered into LLVM-IR {\large \textcircled{\small 7}}, resulting in a \emph{fat binary} that contains the kernel wrapper and launch code for the host and the embedded GPU kernel binaries (PTX in the case of NVIDIA GPUs). This MLIR host code relies on a set of \texttt{mgpu*} function callbacks, such as \texttt{mgpuModuleLoad}, \texttt{mgpuLaunchKernel} and \texttt{mgpuMemAlloc} to provide the GPU-specific functionality. For the Grace-Hopper these functions are implemented using the CUDA API \cite{nickolls:2008:cuda} and added to \exahype{} {\textcircled{\small 8}}. 
Crucially, the \dslhype~generated IR does not need to be changed to target different GPU architectures such as AMD.
Only device-specific pipeline passes and \texttt{mgpu*} functions are required.

The GPU offloading pass implicitly assumes that user functions do not have side-effects and do not read global states. 
If kernels rely on values that are not explicitly passed via arguments, then these values must exist on the accelerator beforehand. With a purely functional mindset, which is the intention of our DSL design explored in Section \ref{sec:dsl_design} this special case can not arise. 
%In the future additional GPU operations in MLIR might explicitly exploit shared memory that is provided by the hardware or GPU drivers, and this would enable us to relax this restriction. 

\section{Benchmark setup}
\label{section:benchmarks}

We assess \dslhype~using the numerical relativity solver ExaGRyPE \cite{Zhang:2025:ExaGRyPE} which is built on top of \exahype~\cite{Reinarz:2020:ExaHyPE}.
It simulates gravitational waves employing CCZ4 compute kernels \cite{Dumbser:2018:CCZ4},
which give us a system of 59 non-linear PDEs written as a non-conservative PDEs plus a source term. 
Furthermore, we also run tests for the Euler equations which describe the evolution of matter.
They involve five equations feeding into a conservative flux.

Both equations are frequently used within fourth-order FD and FV schemes.
This gives us four numerical schemes with different runtime characteristics---from very cheap with low memory footprint (Euler with FV) to very expensive and memory-intense (CCZ4 with FD). 
%Details on the compilation flags we used when building \exahype ~can be seen in appendix \ref{appendix:build}.

\exahype~sits atop of Peano \cite{Weinzierl:2019:Peano}, a framework for PDE solvers on dynamically adaptive Cartesian meshes. 
It is based upon a generalized octree called a spacetree and belongs to the class of cell-wise adaptive AMR codes \cite{Dubey:2021:AMR}. 
The leaves of the spacetree form the compute cells which in turn carry the FD or FV data structures.
Typically, we embed $3 \times 3 \times 3$, $6 \times 6 \times 6$ and $16 \times 16 \times 16$ patches into Peano's mesh.

Peano's mesh traversal automatically bundles compute kernel calls from \exahype~into fused compute kernel calls updating up to $K=8,694$ cells at once \cite{Li:2022:DynamicTaskFusion}. 
These bundles of cells subject to a $\tilde{ \mathcal{K} }_{\Delta t, K}$ kernel call are then shipped to the accelerator, i.e.~\exahype~works with pure kernel offloading, where cell data are not held persistently on the GPU. Data transfer time hence plays a crucial role.

\subsection{Mini-app}

Benchmarking a complex physics code with dynamic AMR, I/O, and massive parallelism yields results influenced by many aspects.
In order to isolate and explore the contribution of this paper, we extract the compute kernels into a stand-alone mini-app (benchmark driver) which enables us to exclusively study kernel runtime and data transfers related to kernel offloading.

Within this mini-app, we always average over 10 runs per measurement while we examine different kernel realizations:

\begin{itemize}
    \item \emph{C++ Kernel} is a plain C++ kernel produced by the simple pretty printer over \dslhype's CST described in Section \ref{section:pipeline}. 
    \item \emph{C++ Kernel with OpenMP} parallelizes the C++ code by adding OpenMP \verb|parallel for| statements to each loop.
      A GPU variant is produced by leveraging OpenMP target offload and adding \verb|teams distribute| statements. 
      Besides the parallelization, the OpenMP variant also collapses loops.       
    \item \emph{MLIR Kernel} is a kernel generated by our \dslhype~MLIR lowering pipeline. 
      This variant can be used in its plain form or in combination with memory flatting (Section~\ref{subsection:pipeline:memory-flattening}).
%      If deployed to a multithreaded environment, both variants are passed through \texttt{scf.parallel} into the MLIR OpenMP \texttt{omp} dialect \cite{xxxx}. 
    \item \emph{GPU MLIR} lowers the IR directly into MLIR's GPU dialect \cite{Webpage:2026:GPU}.
      It always employs memory flattening.
\end{itemize}

\noindent
The plain C++ Kernel serves as a baseline that we measure against.
While it is not aggressively optimized, it exploits certain knowledge about the underlying data such as the fact that the data per cell is organised as continuous AoS memory block.
This is an implementation assumption that we consider to be ``natural'' to exploit for any human programmer, but do not hardcode into \dslhype.

\subsection{Test environment}

GPU benchmarks were run on a Grace-Hopper H200, and CPU benchmarks on an Intel Sapphire Rapid CPU (Intel\textregistered~Xeon\textregistered~Platinum 8480+). 
%The latter hosts two sockets with 56 cores per socket each plus 512GB main memory.
All problem sizes have been tailored such that the H200's 141GB are more than enough to host all data feeding into a computation.
Our compilation pipeline builds upon LLVM/MLIR version 21.1.8. 
All statements on hardware behavior are supported by measurements through Likwid v5.5.1 \cite{treibig:2010:likwid,roehl:2014:likwid}. 

\section{Results}
\label{section:results}

\subsection{CPU single core}
\label{subsection:results:cpu-single-core}

\begin{figure*}[thb]
  \centering
  \includegraphics[width=0.3\linewidth]{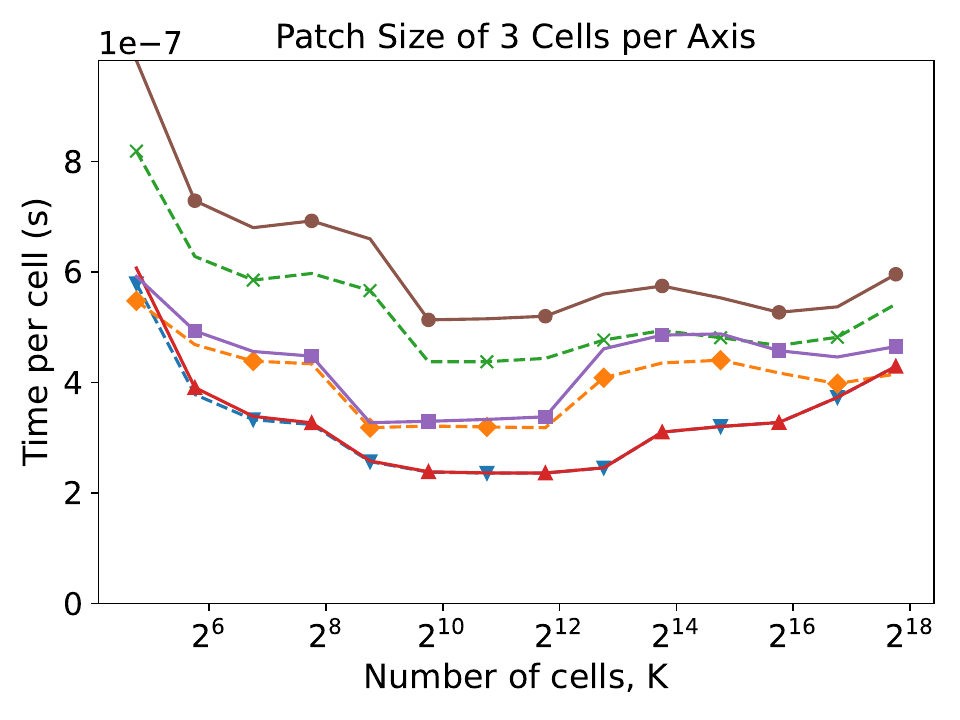}
  \includegraphics[width=0.3\linewidth]{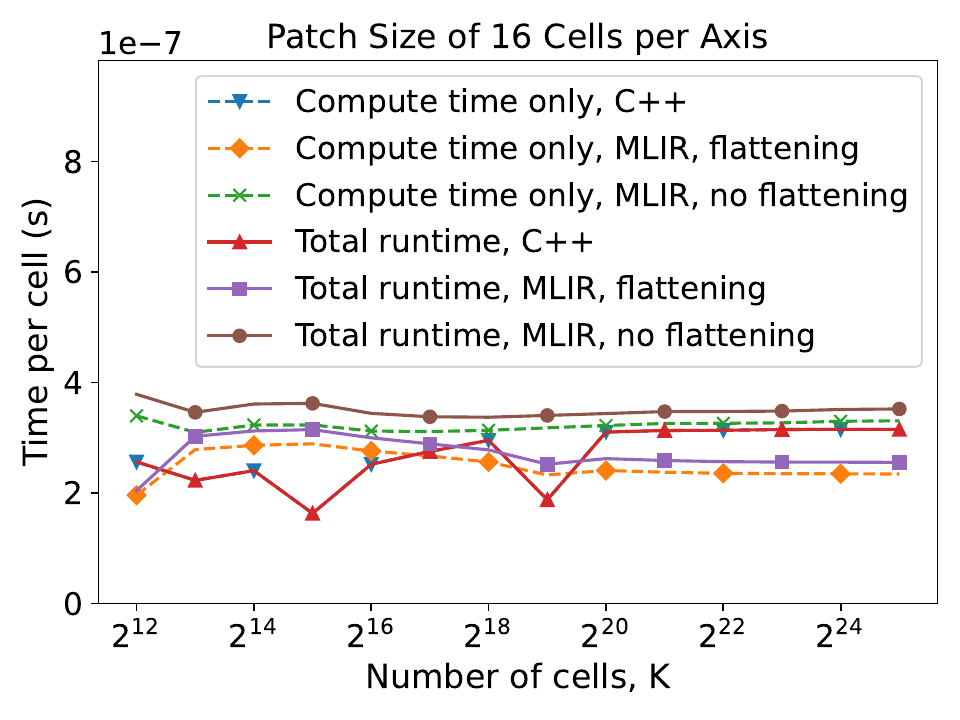}
  \includegraphics[width=0.3\linewidth]{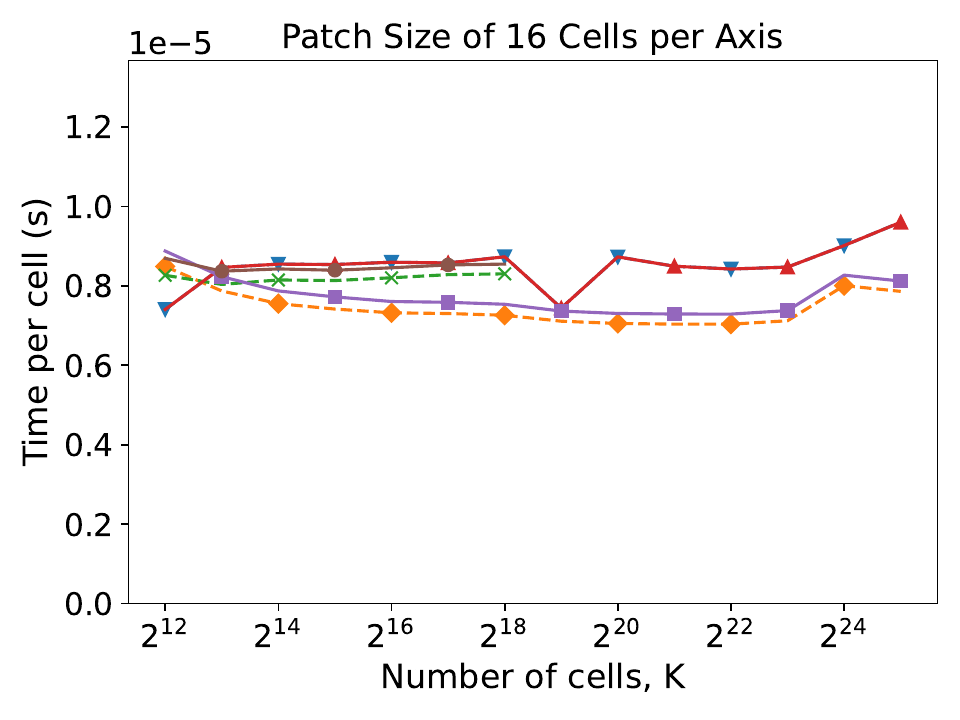}
  \caption{
    Comparison of runtimes per cell for different serial variants with  an FV solver.
    Two Euler examples are presented here (left and middle) and one CCZ4 example (right). The full set of plots are shown in appendices \ref{appendix:raw_data_euler} and \ref{appendix:raw_data_ccz4}.
    \label{fig:results:euler:serial-cpu:sapphire-rapid}
  }
\end{figure*}

%
% What do we do and what do we see:
% - Bigger patches, better performance
% - C++ always exactly the same as total and per kernel time
% - MLIR always slower for small cell counts, eventually always takes over
% - Only for FV do we see a difference of MLIR compute time and MLIR total time, but for both CCZ4 and Euler
%
We first measure single core performance as runtime per grid cell, where timings distinguish pure compute time from total runtime, the later including data preparation, i.e.~any memory rearrangement (Figure~\ref{fig:results:euler:serial-cpu:sapphire-rapid}). 
Larger meshes per cell (patches) lead to higher throughput.
However, batching multiple cells into one compute call has a less uniform impact:
For small patch sizes, increasing $K$ initially pays off, before the runtime starts to rise again with $K$.
For big patch sizes, no such effect is visible.
For CCZ4, the effect is less pronounced than for Euler.
Overall, the MLIR version is faster than the vanilla C++ version throughout all configurations on the CPU.

%
% Explanation and interpretation
%

\emph{Observation: Role of arithmetic load}
A less pronounced impact of settings for CCZ4 is due to the higher arithmetic load within CCZ4's user functions.
CCZ4's user functions already yield such a high compute workload that the performance character is not that sensible to the loops around them. 
For Euler, the context in which user functions are called makes more of a difference.
A similar reasoning holds for higher-order FD, where calculations imply that other steps such as data rearrangements make a smaller relative runtime contribution.

% \emph{Observation: Collapsing}
% Loop collapsing as offered through OpenMP is an absolute key ingredients to yield % high performance.
% MLIR's OpenMP dialect does not provide this feature to the best of our knowledge.
% Therefore, the MLIR version struggles to compete with its vanilla C++ cousin.

%\emph{Observation: Vectorisation}
%The memory flattening and aggressive inlining facilitates higher vectorisation.
%It eventually allows the MLIR-generated kernel to outperform is native cousin once the workload becomes reasonably large.
%Initially however, the flattening introduced an overhead to create and befill the temporary memory which manifests in a higher memory transfer footprint.

\emph{Observation: Data size impact}
Both the MLIR and the vanilla C++ versions run through the kernel algorithm step by step, providing a plain realization of expression \eqref{equation:challenge:operator-composition}.
Therefore, a large $K$ induces cache capacity and Translation Lookaside Buffer (D-TLB) misses between individual steps.
Our kernels implement aggressive loop fission, even though the loop over $K$ could be factored out of the kernel, i.e.~fused between different steps. 
%We use hardware performance counters on the Intel Sapphire Rapids chip to verify these predictions.
%Such a loop re-orchestration yields higher performance \cite{Loi:2024:SYCL}, yet is not integrated into our compiler pipeline at the moment.

To validate the statements on memory access characteristics, we focus on a single case with patch size of 3 and $K=2^{14}$ patches. 
For the CCZ4 kernels, the data volume from main memory for the MLIR implementation without memory flattening is approximately 114\% of the native C++ representation. 
With memory flattening this reduces to  approximately 102.5\%. 
We likewise see pressure on the D-TLB, as there is an almost 7\% increase in cycles spent in D-TLB load misses for the MLIR implementation without memory flattening relative to the native C++. 

The most stark representation of the increased memory and cache management in the MLIR implementation results from the Top-down Microarchitecture Analysis (TMA) counters \cite{yasin:2014:tma}.
They show that MLIR without memory flattening increases the L2 cache memory bound metric by almost 57\%, though again with memory flattening this difference drops to 2\% relative to the C++ implementation. 
The MLIR implementation therefore consistently appears to have less efficient memory management than the native C++ kernels, but the inclusion of the memory flattening pass improves this significantly.

\subsection{GPU results}
\label{subsection:results:gpu}

\begin{figure}[htb]
  \centering
  \includegraphics[width=0.66\linewidth]{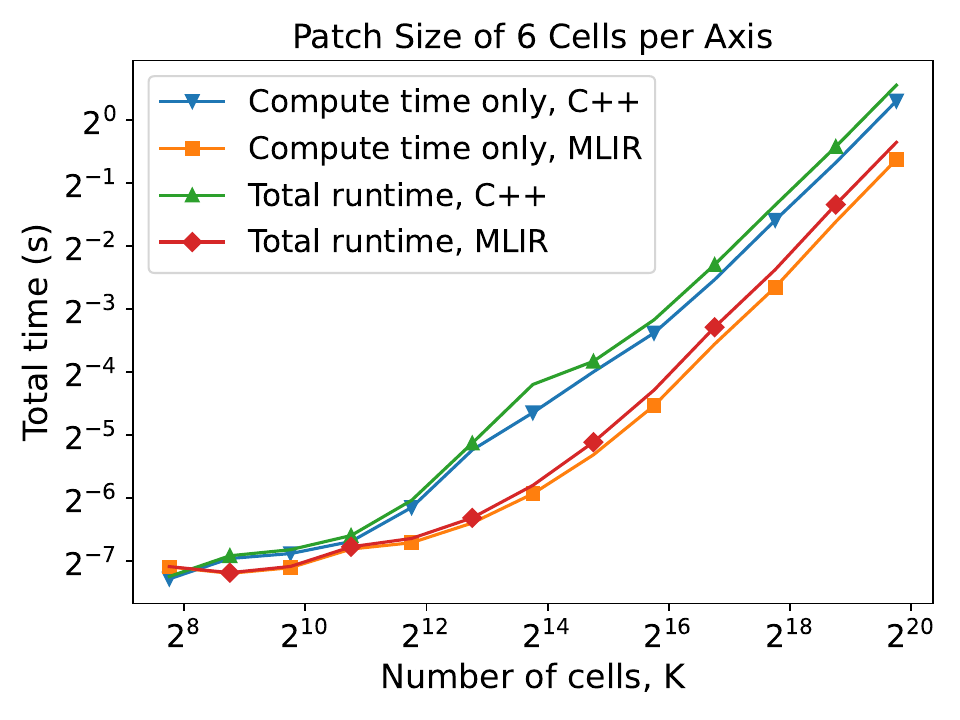}
  \includegraphics[width=0.66\linewidth]{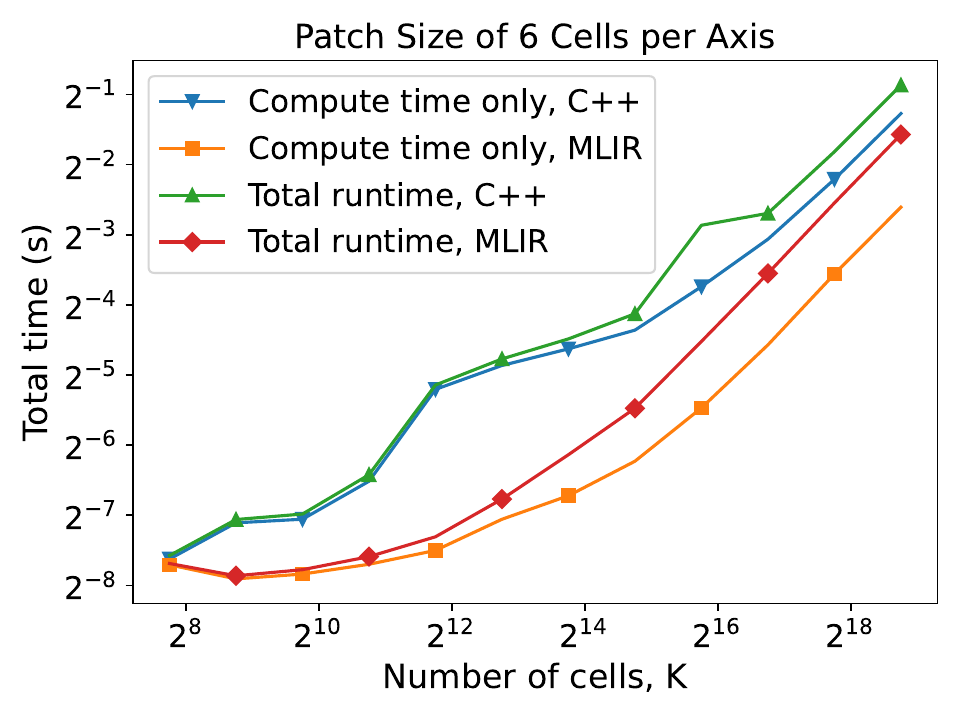}
  \caption{
    Runtime over all cells of CCZ4 on the Hopper GPU architecture with low order FV (top) and higher-order FD (bottom). 
    \label{fig:results:euler:gpu:hopper}
  }
\end{figure}

%
% Whta we do and what we see
%

We continue to compare performance of the CCZ4 benchmark on GPUs using \dslhype{} to the C++ generated code employing OpenMP offloading (Figure~\ref{fig:results:euler:gpu:hopper}). 
The comparison of \emph{compute time} against \emph{total runtime} shows a significant difference, since GPU offloading imposes overheads that add significantly to the total runtime, notably for the higher-order scheme, where we have to work with significant halo layers.
Different to the CPU version, employing \dslhype~nevertheless pays off uniformly,  and the total compute time starts to grow linearly with the compute workload after a burn-in phase for small $K$.

\emph{Observation: Compute saturation}
The numerical schemes expressed through \dslhype~are well-suited for a streaming multiprocessor, as they outperform their CPU counterpart (comparison data not shown).
Counterintuitively, the GPU calculates higher-order schemes faster than a low-order scheme.
However, a certain number of cells is required to saturate the GPU's compute capabilities and to allow the GPU to unfold its potential.
Experimenting with different kernel variants produced through \dslhype~might hence suggest that only higher-order cell kernels are offloaded to GPUs and the offloading is triggered if and only $K$ is sufficiently large.

\emph{Observation: Overheads}
Kernel launches, data re-organization and data migration overheads double the higher-order's runtime, and these are particularly significant for the MLIR version. Flattening and data transfer scales almost linearly with the data footprint, as this is pure data movement overhead, and higher-order FD schemes require larger overlaps (halo layers) than the low-order FV. It is a subject of future work to explore to which degree GPU streaming and multithreading on the host can hide the data reordering and transfer overheads.

\subsection{Lines of Code}

\begin{figure}[htb]
    \centering
    {\footnotesize
    \begin{tabular}{c|c|c|c}
         & DSL & Handwritten CPU & Handwritten GPU \\
         \hline
         FV& 85 & 660 & 1124 \\
         FD4& 73 & 752 & Not Implemented \\
    \end{tabular}
    }
  \caption{
    Numbers of lines of code required to implement 3D solvers with and without the DSL.
    These numbers count only core numerical loops.
    They do no include user-defined lines from Euler or CCZ4.
    \label{fig:loc}
  }  
\end{figure}

Using the DSL significantly reduces the effort required to implement a new solver:
the developer only needs to implement one version independent of the dimensions $d$, 
the numerical scheme is written as numerical specification and the optimization is deployed to the compiler,
and the toolchain generates CPU and GPU outputs automatically. 
For a genuine C++ implementation, all variants have to be maintained manually,
while is is up to the programmer to ensure that all user-defined functions hosting the actual physics are made available for the GPU, too.
\dslhype{} hence reduces the lines of code required to write down our FV and FD4 solvers (Table \ref{fig:loc}).

%Our FV solver previously required ~1000 LOC providing utility functions, then for each version of the solver there was a file in the range of 500-900 LOC using these functions. There were versions for each combination of concurrency, loop structure, and in the case of GPU usage, backend API. In contrast, with DSLHyPE we require only a version for each number of dimensions, each ~90 LOC. We have similar reductions for our FD4 solver.

\section{Evaluation and thoughts on appropriate optimization passes}
\label{section:optimisation}

Whilst our experiments in Section \ref{section:results} demonstrate that MLIR compares favorably against the C++ baseline, in undertaking these measurements we have made several observations that help us to identify opportunities for further optimizations. 
On the CPU in particular, well-chosen loop transformations can improve cache usage and, combined with instruction-level parallelism considerations, boost runtime \cite{Wolf:1996:CombiningLoopTransformations}.
This section explores which additional passes might be particularly beneficial based upon the behavior of \dslhype{} and the performance properties observed.
However, it focuses on \dslhype-specific optimizations which are not yet available within the set of canonical MLIR optimization passes.  
%This list is neither comprehensive nor does it imply that other, traditional MLIR passes such as loop reordering or blocking lack performance impact.

\subsection{Kernel concatenation and fusion}

\dslhype's front-end maps the projections of data blocks from the source code (Figure~\ref{figure:compute-kernel}) onto a sequence of $S$ kernel invocations.
Since kernel invocations are expensive, it is natural to fuse the concatenated kernels into one.
As individual kernels have different iteration ranges, a fused kernel might have to run over the maximum of all involved ranges and mask out indices at certain steps.
Certain subranges (warps) might also require warp-local barriers.
We discuss these techniques in \cite{Loi:2024:SYCL}.

Fusing the loops hosted in long, complex kernels is not without problems on GPUs, as it tends to increase register pressure and amplify memory access challenges.
High throughput hence may only be achievable if such optimizations are carefully evaluated and applied \cite{Wang:2025:AdaptiveFusion}.
Register occupation metrics might help to guide such transformations.

\subsection{AoS to SoA conversion}

Vectorization across multiple user functions is often key to high performance \cite{Pekkila:2025:TuningStencils}, yet does not happen automatically in \dslhype's vanilla implementation.
One reason is the inherent realization of user functions over AoS.
A natural optimization candidate is therefore to reorder the input data into SoA, which can be added to the preamble and epilogue operations from \eqref{equation:challenge:gather-scatter}.

As the user code is automatically inlined into the created MLIR kernel's IR, such a memory conversion would be completely hidden from the user, while experiments with other algorithm types suggest that the transformation can yield massive performance improvements \cite{Radtke:2025:SoAtoAoS}.
However, the reordering adds additional scattering to the two memory-bound preparation and epilogue steps.
Appropriate pipelining and loop fusion (e.g.~overlapping operations in $\mathcal{K}_{\Delta t}^{(0)}$ and $\mathcal{K}_{\Delta t}^{(1)}$) might hence be necessary.
Along these lines, it might also be beneficial to move the rearranging from the host code onto the GPU.

\subsection{Collapsing and unification of temporary data structures}

\dslhype's front-end creates one temporary data structure per step, unless data blocks are explicitly tied to input or output data.
This introduces several dynamic memory allocations per kernel call and can lead to a substantial temporary memory footprint.
Dynamic memory allocations are expensive---particularly on GPUs \cite{Wille:2023:GPUOffloading}.

While memory pooling is now a state-of-the-art technique \cite{Wille:2023:GPUOffloading,Qian:2023:MemoryPool}, \dslhype's design makes it possible, in theory, to analyze the total memory footprint ahead of time and allocate one large scratchpad per kernel invocation.
Alternatively, a compiler optimization pass could reorder steps and explicitly recycle temporary memory by passing it from one step to the next.

\subsection{Kernel parallelization and orchestration}

Mapping the steps from \eqref{equation:challenge:gather-scatter} onto a sequence of nested loops ignores that these steps could potentially run in parallel.
Once loops are fused, a compiler or scheduler might implicitly parallelize some operations.
Loop permutation could further improve runtimes.
We have studied the corresponding techniques and labelled them as horizontal and vertical parallelisation \cite{Loi:2024:SYCL}.
It is clear that our manual arrangements (written in SYCL) could be moved into compiler optimization passes,
but which combination of different approaches within the vast search space to pick remains an open problem.

As alternative to loop reordering, one can identify such parallelism explicitly and express the potential concurrency directly.
Modern (GPU) programming languages such as OpenMP or SYCL allow compute steps to be modeled as directed, acyclic graphs.
The runtime can then exploit any arising parallelism by deploying individual graph nodes (mini-kernels) onto separate warps or cores.
While we have not found this approach beneficial when applied throughout an entire compute kernel \cite{Loi:2024:SYCL}, such parallelism might pay off for individual steps of especially expensive kernels.

\subsection{Tensor formulations and explicit linear algebra}

Stencil steps in \dslhype~span a large, potentially sparse linear equation system applied over the input data block.
Sequences of stencil steps hence form sequences of matrix-vector multiplications.
For linear kernels, it is known that all user functions can be evaluated once at the kernel start, with the outcomes then combined through a series of tensor contractions \cite{Uphoff:2020:YaTeTo}.

It is hence a natural choice to construct compiler optimization passes that identify such (partial) kernels and map the arising operators onto linear algebra calls.
Indeed, MLIR offers a bespoke dialect for this purpose.
Relying on calls such as BLAS means that all hardware-aware optimization knowledge feeds immediately into the realization of the \dslhype~kernel.

\section{Related work}
\label{section:related-work}

Approaches to generating PDE compute kernels differ mainly in \emph{where} they draw the boundary between the user's problem description and the compiler's responsibility,
and to which degree numerical and problem specification are exposed to the user. 
\dslhype{} keeps the physics and the numerics completely seperate, written in different languages, and fuses them late, inside the compiler,
while all optimization of the merger is deployed to the compiler stack. 
 
Frameworks such as FEniCS~\cite{fenics} and Devito~\cite{lange2016devito} represent the entire problem including the equations, discretization and schedule, in one mathematical language and encode the numerical and HPC expertise within their translation recipes, ultimately emitting complete applications. \dslhype{} instead treats the PDE terms of~(\ref{equation:introduction:PDE}) as black boxes and emits individual kernels that link into an existing engine. 
This matters for our astrophysical setups, where a single term can couple a large number of nonlinear PDEs and where a fully symbolic formulation is impractical or not desireable:
Users retain validated, pre-existing term implementations.
 
Syntactically, PyStencils~\cite{pystencils} is closest to our front-end, generating CPU and GPU functions from Python stencil definitions. Its focus on classic FD/FV stencils, however, excludes the whole-array matrix operators \dslhype{}'s \texttt{DataBlock}s provide, which are essential for DG. ExaStencils~\cite{exastencils}, Snowflake~\cite{snowflake} and MSC~\cite{msc} likewise contribute sophisticated dependence analyses and loop transformations, but operate as source-to-source systems and they must therefore implement and maintain these optimizations internally. \dslhype{} deliberately outsources this where, by lowering into MLIR, it inherits the community's optimization and GPU lowering passes, and contributes only the bespoke passes (physics inlining, memory linearization, memory-space rewriting) that our bilingual design requires.
 
Toolboxes such as YATeTo~\cite{yateto} evaluate all PDE terms once and combine the results through staged tensor contractions, unlocking dense linear algebra optimizations for linear kernels. \dslhype{} targets the complementary regime where nonlinear kernels in which term evaluations sit deep inside the calculation flow, are invoked repeatedly, and consume intermediate results computed on the fly.
 
Triton~\cite{triton} defines GPU kernels via decorated Python functions with block-level parallelism, while POM~\cite{pom} extracts polyhedral semantics into MLIR for FPGA accelerators. Both are monolingual. To the best of our knowledge, the \emph{bilingual} integration of native user code with a Python numerics description at the MLIR level which is realized here via Polygeist~\cite{polygeist} is novel. The enforced cross-language inlining it entails is precisely what exposes the whole-kernel optimization opportunities of Section~\ref{section:optimisation} to the compiler, and crucially these are opportunities that per-language pipelines cannot see.
\section{Summary and outlook}
\label{section:conclusion}

We introduce a domain-specific language and corresponding compiler, \dslhype, for compute kernels in our \exahype~engine.
Since the compiler builds on MLIR, its implementation is lightweight and focuses solely on the front-end parts of the translation.
\dslhype's guiding design principle is to separate user code (physics) from numerical schemes: users provide their domain knowledge in native C or C++, while the numerics are modeled in a stencil-like language.
Leveraging Polygeist \cite{Moses:2021:Polygeist,Webpage:2026:Polygeist}, the compiler integrates these two languages, interplaying with other, standard MLIR transformations and optimizations.

Many compiler pipelines favor schemes where the code is first expanded to increase the number of potential parallelization, then parallelized, and finally optimized at the level of the individual (parallel) calculations \cite{Thomas:2026:AutomaticVectoriseCompilers}.
\dslhype~fits this mindset: integrating two languages---which can be read as enforced inlining---amounts to exactly such an increase in concurrency.
We therefore think there is great potential in \dslhype's approach.

\dslhype~is used in production by our astrophysics code ExaGRyPE, built on top of \exahype.
Here, we rely on a manual composition of MLIR's optimization passes.
This yields sufficiently efficient code, but does not yet unlock the full potential of a compiler-/DSL-based kernel encoding.
The major gains \dslhype~introduces so far are a strict separation of concerns, support for rapid prototyping of novel numerics, cleaner code, and seamless support for both CPU and GPU kernels.
Efficiency gains are left to future work, though we already sketch some key directions of research.

Notably heuristics, i.e.~cost models \cite{Wang:2025:AdaptiveFusion,Wolf:1996:CombiningLoopTransformations}, could guide this optimization.
As is typical for compiler optimizations, whether and how much a given optimization helps depends on the system, problem configuration, and various other parameters.
Deriving appropriate decision rules is beyond the scope of this work, though an exhaustive search of the configuration space could offer useful, if costly, insight.
Alternatively, tuning frameworks can help choose good parameters \cite{StencilAutoTuning,GeST,PerformanceTuningFramework}, and there is scope for machine learning \cite{Wang:2025:AdaptiveFusion}.

\bibliographystyle{IEEEtran}
\bibliography{paper}

\newpage
\appendix 
\subsection{Reproducibility}

\dslhype~is freely available as part of \exahype. 
\exahype's branch \texttt{2026DSLHyPE} contains all \dslhype~flavours used in the present text.
The code description can be found at \url{http://www.peano-framework.org}, and comes along with tutorials and source code description.
The appropriate version of the code can be cloned via

{\footnotesize
\begin{lstlisting}
git clone --depth 1 --branch 2026DSLHyPE
    https://gitlab.lrz.de/hpcsoftware/Peano.git
\end{lstlisting}
}

\subsection{Download and build}
\label{appendix:build}

The following instructions for compiling the kernel benchmarks assume that GNU Autotools, CMake, and an LLVM build including Clang and MLIR are already installed and added to the PATH variable.
The code snippet

{\footnotesize
\begin{lstlisting}
python3 -m venv env
source env/bin/activate
pip install -r ${PEANO_DIR}/requirements.txt
\end{lstlisting}
}

\noindent
installs all required Python modules.
It assumes that the path \lstinline!PEANO_DIR! is well-defined.

\subsubsection{Building \exahype/ExaGRyPE}

\exahype~\cite{Reinarz:2020:ExaHyPE} and ExaGRyPE are all contained and built from Peano's repository.
It holds not only the adaptive meshing core but also the software packages built on top of this particular mesh.
Its Autotools build system is initialized through

{\footnotesize
\begin{lstlisting}[breaklines=true]
libtoolize; aclocal; autoconf; autoheader; cp src/config.h.in .
automake --add-missing
\end{lstlisting}
}

\noindent
and further configured with 

{\footnotesize
\begin{lstlisting}[breaklines=true]
./configure CC=clang CXX=clang++ --with-multithreading=omp CXXFLAGS="-O3 -std=c++20 -fopenmp -march=native -mtune=native -fomit-frame-pointer -fno-fast-math -frounding-math -Wno-attributes=clang:: -w -g -stdlib=libc++" LDFLAGS="-fopenmp -stdlib=libc++" --enable-loadbalancing --enable-exahype --enable-particles --enable-blockstructured --enable-finiteelements
\end{lstlisting}}

for CPU runs, and for Grace-Hopper runs with

{\footnotesize
\begin{lstlisting}[breaklines=true]
./configure CC=clang CXX=clang++ --with-multithreading=omp --with-gpu=omp CXXFLAGS="-O3 -std=c++20 -fopenmp -march=native -mtune=native -fomit-frame-pointer -fno-fast-math -frounding-math -Wno-attributes=clang:: -w -g -stdlib=libstdc++ -fopenmp-targets=nvptx64-nvidia-cuda --offload-arch=sm_90" LDFLAGS="-fopenmp -stdlib=libstdc++ -fopenmp-targets=nvptx64-nvidia-cuda --offload-arch=sm_90" --enable-loadbalancing --enable-exahype --enable-particles --enable-blockstructured --enable-finiteelements
\end{lstlisting}
}

\noindent
If you do not want to use the standard library that comes with LLVM, remove the \lstinline!stdlib=libc++! lines.
We compile with a simple \lstinline!make! and obtain all libraries forming a complete \exahype~installation.

\subsubsection{Compiling the LLVM software stack}

We next build the \verb|dslhype-opt| tool

{\footnotesize
\begin{lstlisting}
cd {PEANO_DIR}/src/exahype2/mlir
mkdir build && cd build
cmake -G "Unix Makefiles" .. \
  -DMLIR_DIR=${LLVM_INSTALL_DIRECTORY}/lib/cmake/mlir
cmake --build . 
\end{lstlisting}
}

\noindent
and then add the tool to the path:

{\footnotesize
\begin{lstlisting}
export PATH=${PEANO_DIR}/src/exahype2/mlir/build/bin:\
  $PATH
\end{lstlisting}
}

\subsubsection{Create the benchmarks}

To build the \dslhype~benchmarks, we invoke

{\footnotesize
\begin{lstlisting}
cd {PEANO_DIR}/benchmarks/exahype2
cd ccz4/kernel-benchmarks
export PYTHONPATH=../../../../python:\
  ../../../../applications/exahype2/ccz4
python kernel-benchmarks-fv-rusanov-dsl.py \
  --enable-mlir -cpu 
\end{lstlisting}
}

\noindent
which picks up the library configurations of \exahype, runs the Python script wrapping around the miniapps,
invokes the custom LLVM build and eventually executes the benchmarks.
Passing \lstinline!--help! as argument into the benchmarking script provides further guidance on benchmark variants.
An analogous benchmark is shipped for Euler.

%\subsection{Performance analysis}

%We evaluate our kernel variants with various tools.
%The analysis is inspired the methodology proposed by the SHAREing project (\url{https://shareing-dri.github.io}),
%where we focus on the intra-node and the core-level analysis, as well as the GPU metrics.

%\subsubsection{Program phases}

%We first localise our measurements.
%That is, we separate the gather and scatter steps from the actual compute.
%Gather and scatter are intrinsically memory-bound, as they solely move data around.

\subsection{Limitations}

Our proposed solution should, in theory, work for any application following \eqref{equation:introduction:PDE}.
In practice, we did face some limitations:

\begin{enumerate}
  \item Not all of C's mathematics functions are currently supported fo the GPU lowering. 
    It struggles in particular with functions like \texttt{fmax} and \texttt{fmin}.
    Replacing these functions if if-then-else branches eliminates this lowering issue.
  \item The \texttt{mlir} inlining pass struggles with extensive dynamic memory allocations on the stack.
    Replacing these allocations with genuine local variables, i.e.~avoiding any dynamic allocation, helps to eliminate this issue.
    We have observed similar behavior with our vanilla OpenMP version as produced by the pretty printer.
\end{enumerate}
\onecolumn
\subsection{Raw data: Euler}
\label{appendix:raw_data_euler}
\subsubsection{Serial CPU}
\begin{figure}[!htb]
  \centering
  \includegraphics[width=0.3\linewidth]{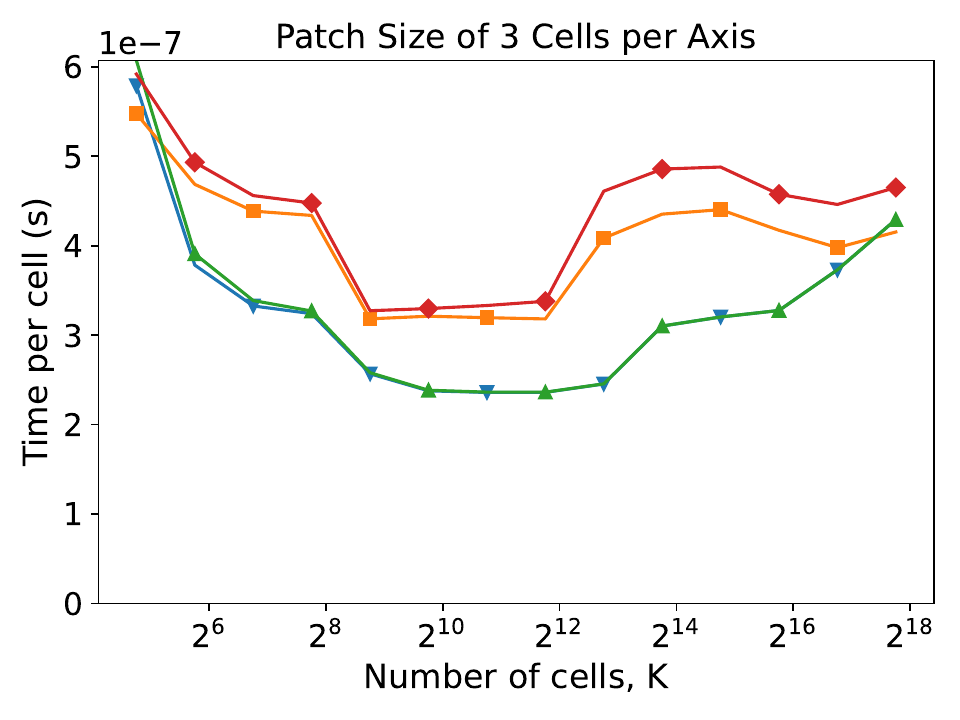}
  \includegraphics[width=0.3\linewidth]{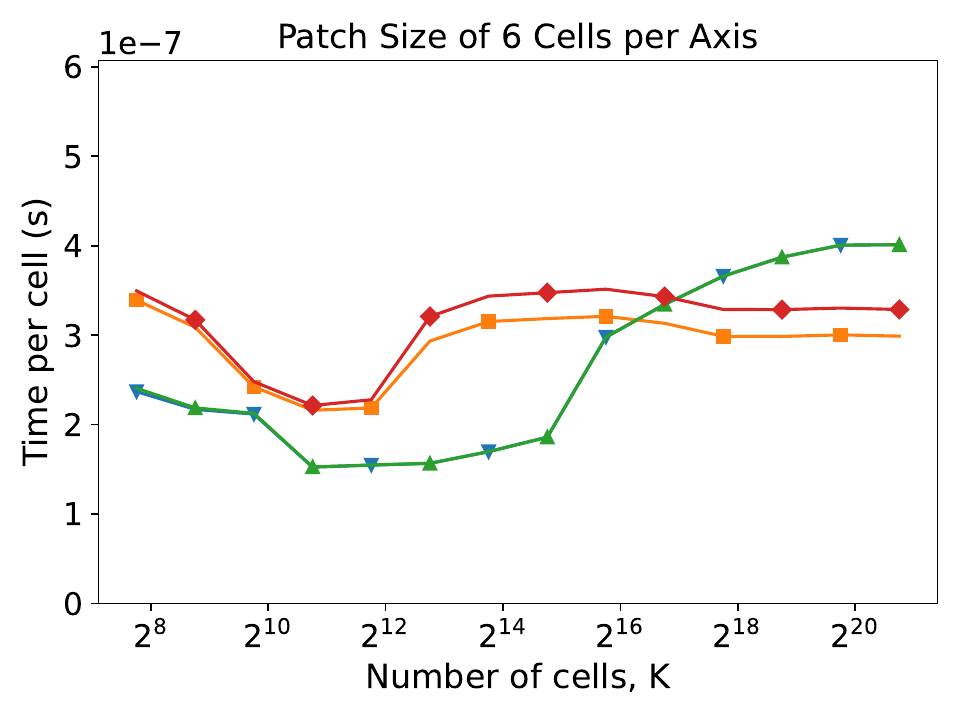}
  \includegraphics[width=0.3\linewidth]{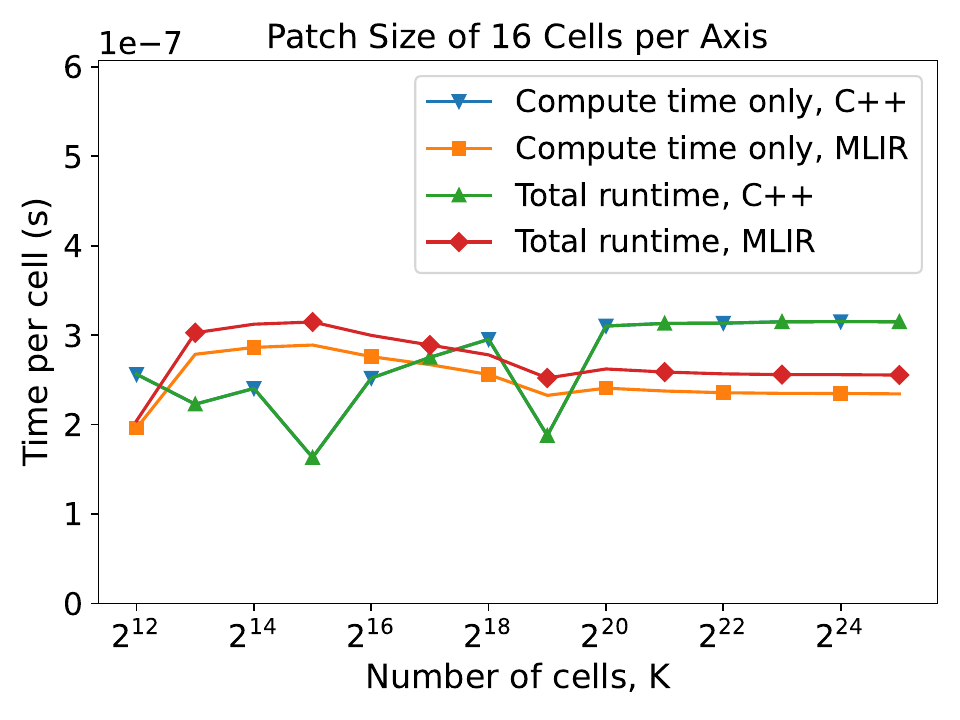}
  \caption{
    Runtime per cell of Euler with Finite Volumes using an FV solver.
}\label{fig:appendix:plots:euler:rusanov:cpu_serial:sapphire_rapid:normalised}
\end{figure}

\begin{figure}[!htb]
    \centering
    \includegraphics[width=0.3\linewidth]{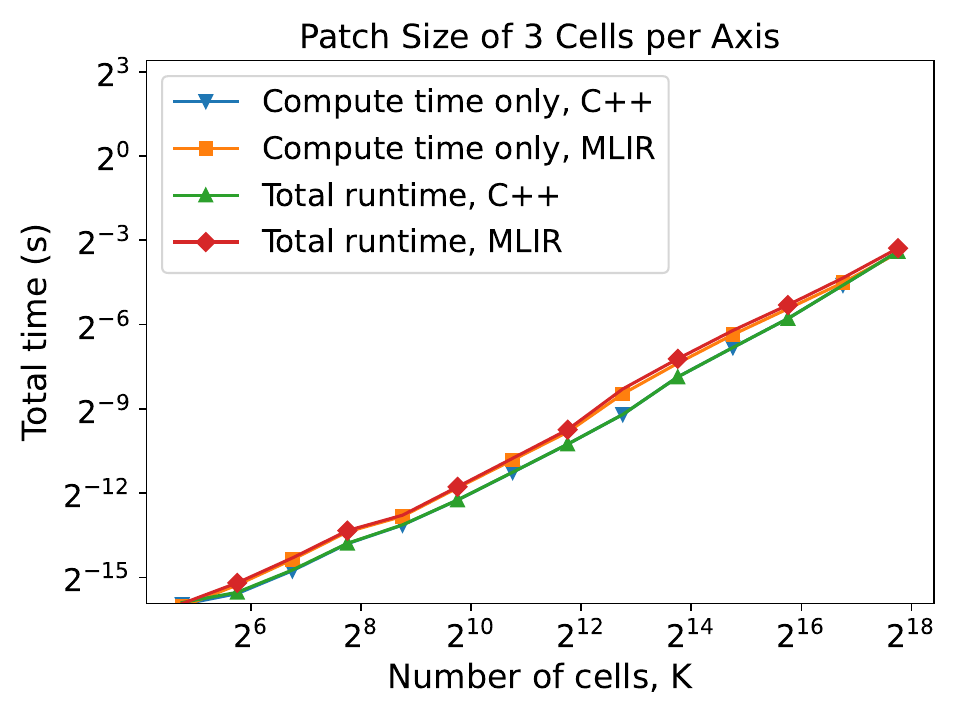}
    \includegraphics[width=0.3\linewidth]{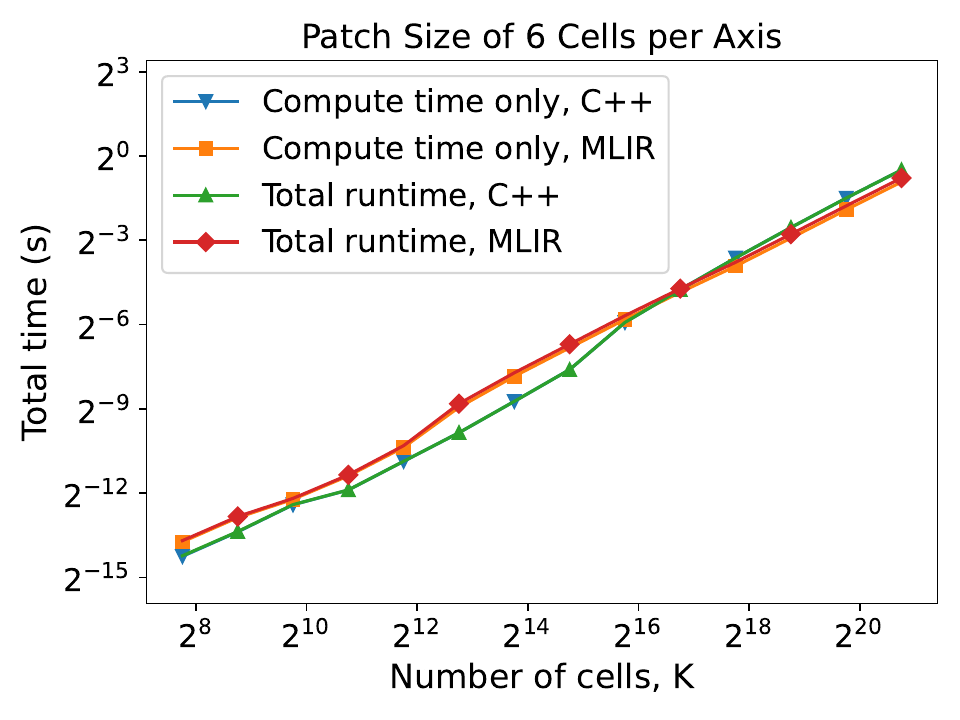}
    \includegraphics[width=0.3\linewidth]{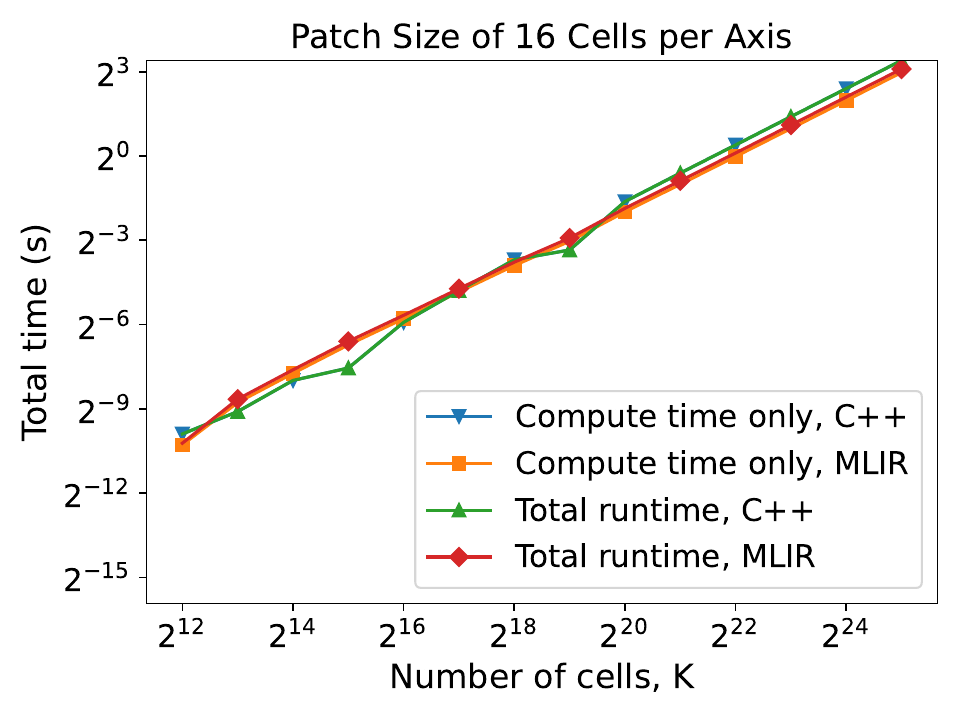}
    \caption{Runtime over all cells of Euler with an FV solver.}
    \label{fig:appendix:plots:euler:rusanov:cpu_serial:sapphire_rapid:log}
\end{figure}

\begin{figure}[!htb]
    \centering
    \includegraphics[width=0.4\linewidth]{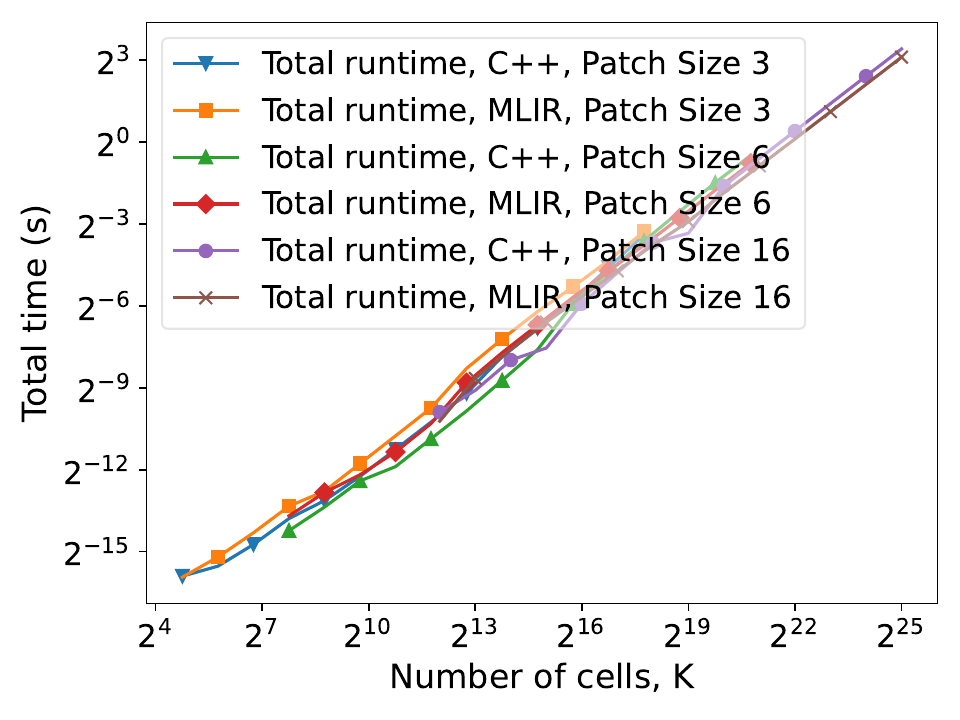}
    \includegraphics[width=0.4\linewidth]{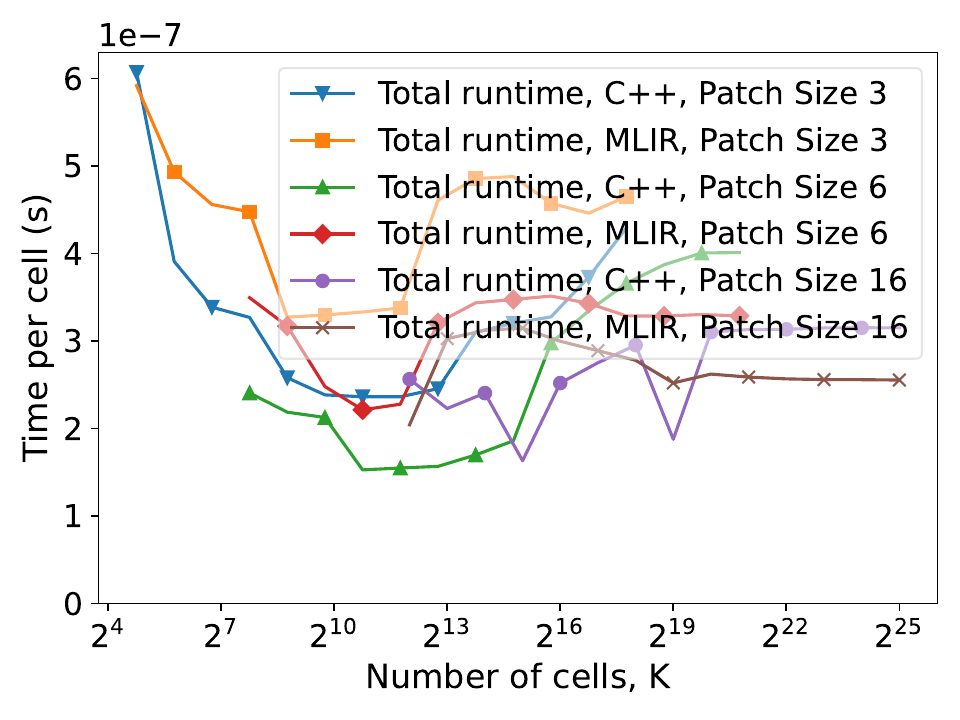}
    \caption{Total runtime for Euler with an FV solver with different patch sizes. Left: Total runtime over all cells, right: runtime per cell.}
    \label{fig:appendix:plots:euler:rusanov:cpu_serial:sapphire_rapid:all}
\end{figure}

\begin{figure}[!htb]
  \centering
  \includegraphics[width=0.3\linewidth]{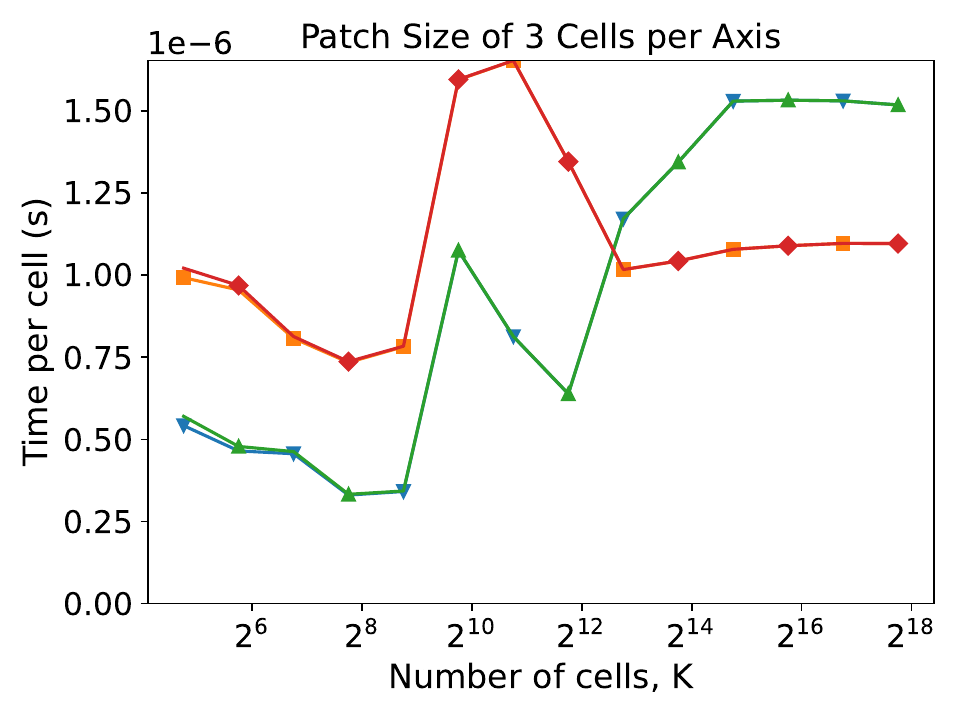}
  \includegraphics[width=0.3\linewidth]{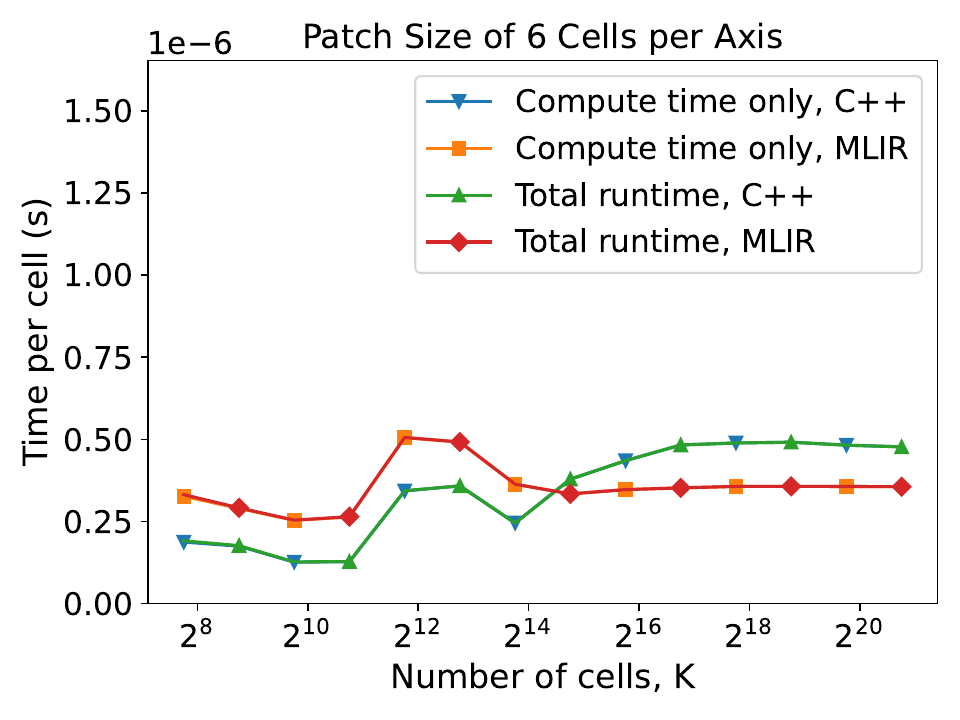}
  \includegraphics[width=0.3\linewidth]{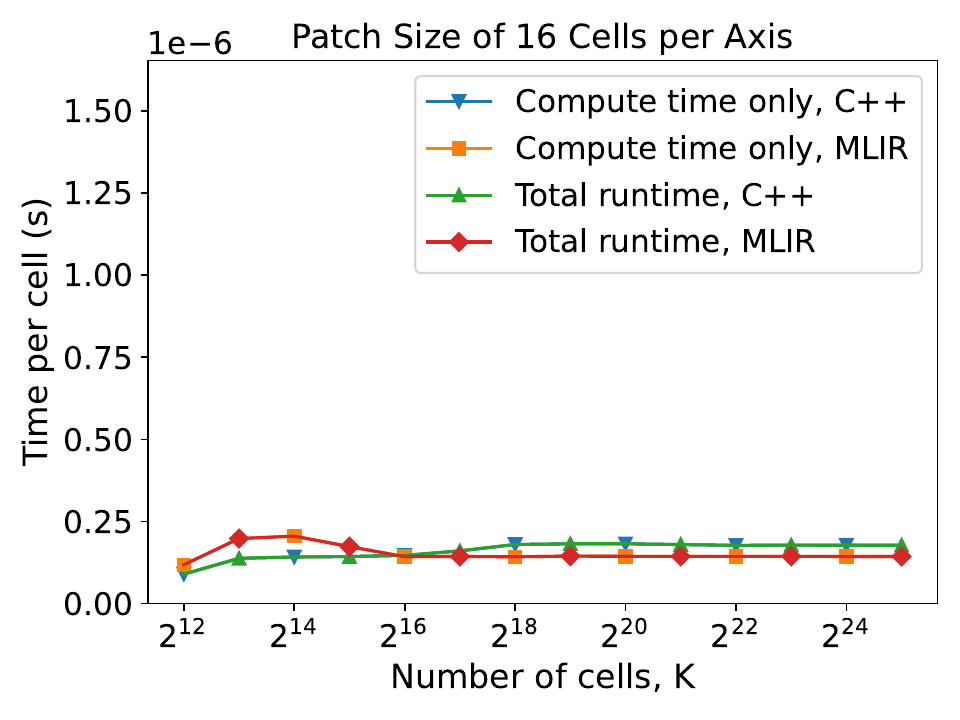}
  \caption{
    Runtime per cell of Euler with Finite Volumes using an FD4 solver. \label{fig:appendix:plots:euler:fd4:cpu_serial:sapphire_rapid:normalised}
  }
\end{figure}

\begin{figure}[!htb]
    \centering
    \includegraphics[width=0.3\linewidth]{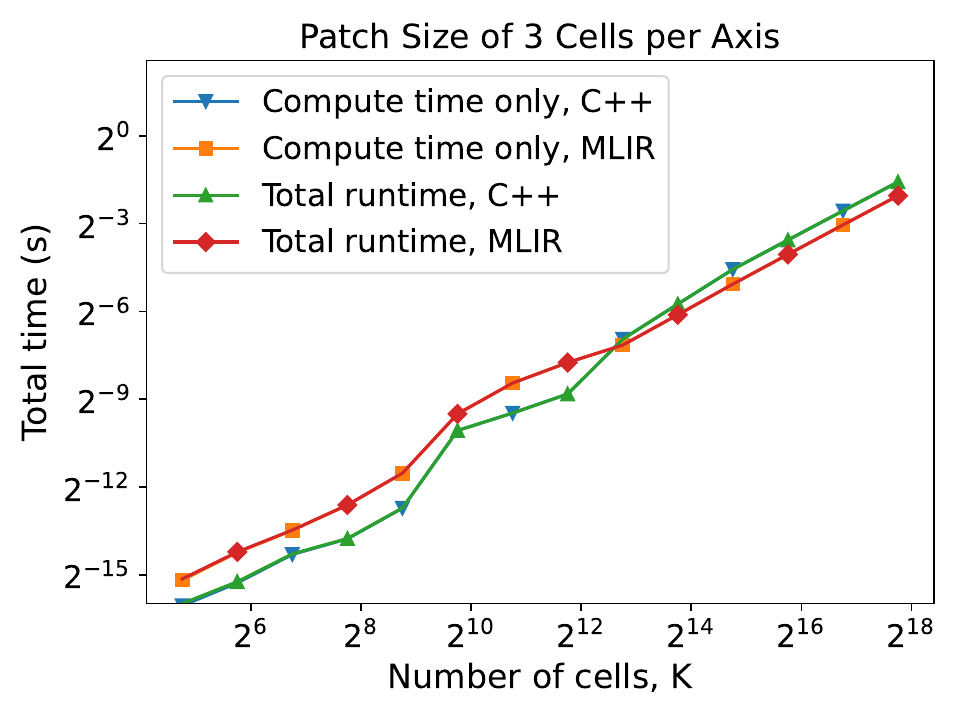}
    \includegraphics[width=0.3\linewidth]{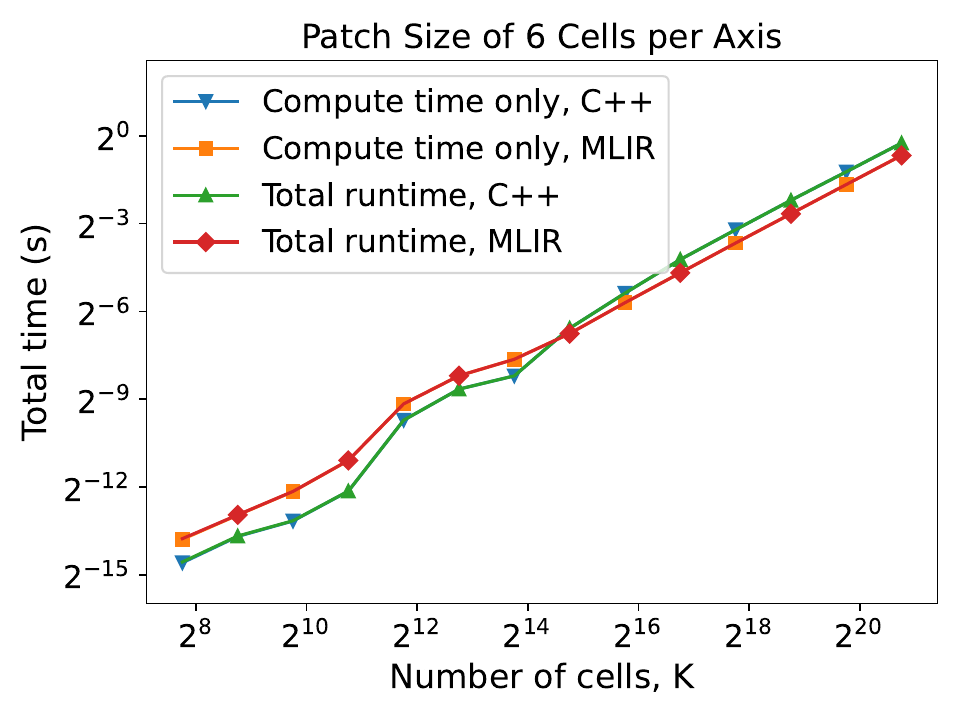}
    \includegraphics[width=0.3\linewidth]{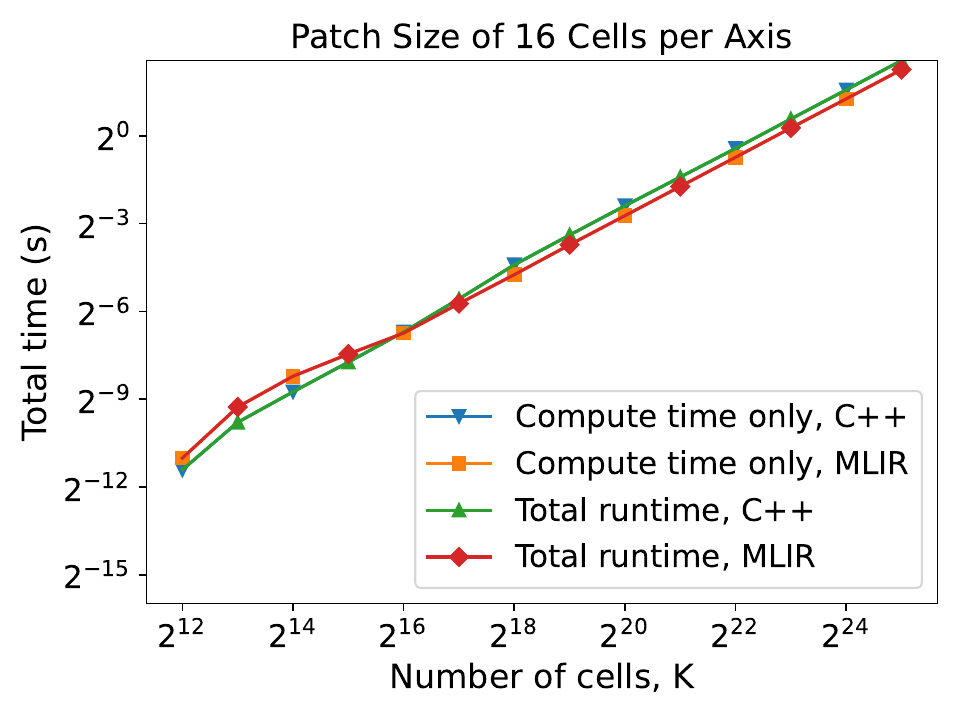}
    \caption{Runtime over all cells of Euler with Finite Volumes using an FD4 solver.}
    \label{fig:appendix:plots:euler:fd4:cpu_serial:sapphire_rapid:log}
\end{figure}

\begin{figure}[!htb]
    \centering
    \includegraphics[width=0.4\linewidth]{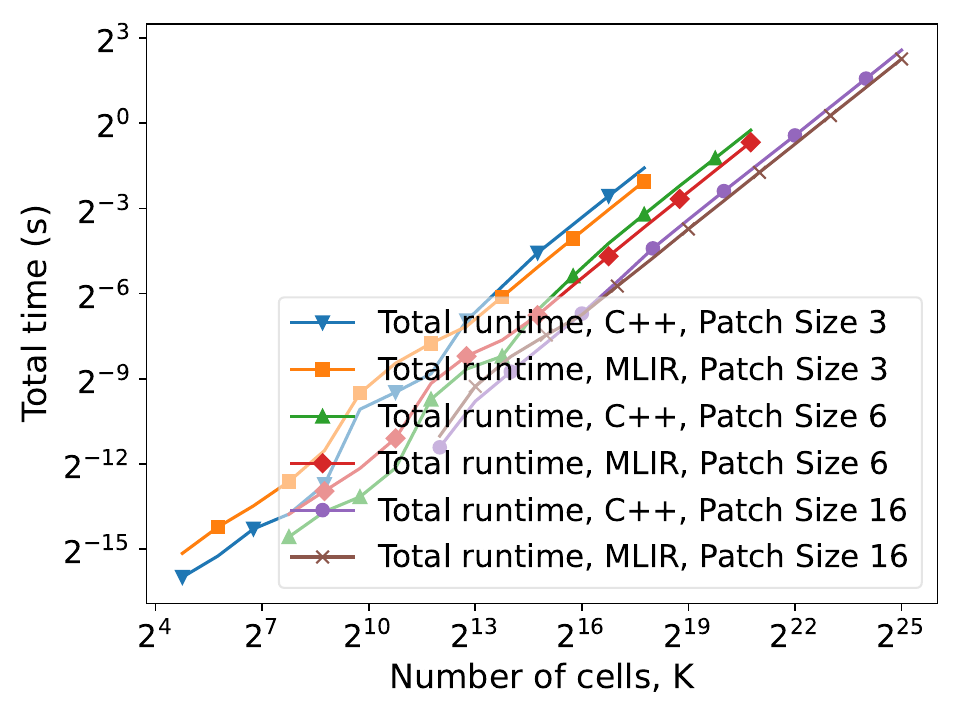}
    \includegraphics[width=0.4\linewidth]{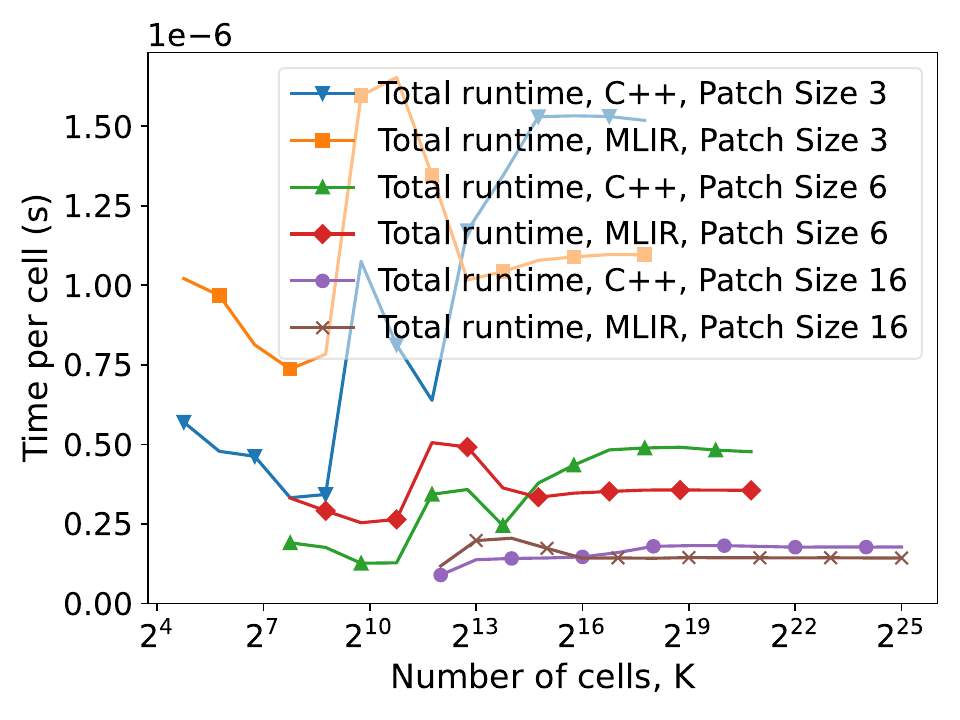}
    \caption{Total runtime for Euler with an FD4 solver with different patch sizes. Left: Total runtime over all cells, right: runtime per cell.}
    \label{fig:appendix:plots:euler:fd4:cpu_serial:sapphire_rapid:all}
\end{figure}

\begin{figure}[!htb]
    \centering
    \includegraphics[width=0.3\linewidth]{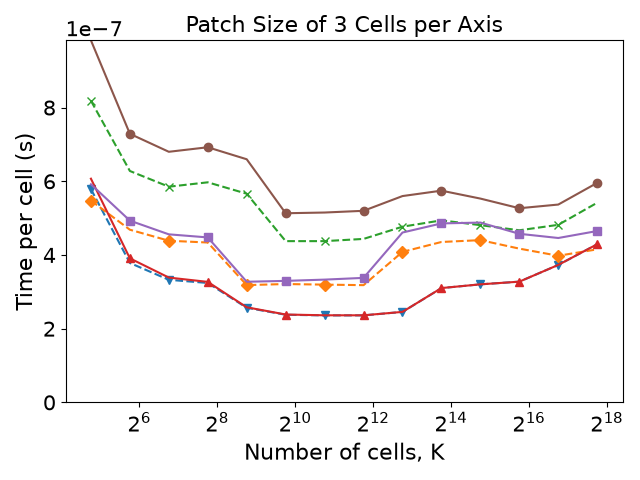}
    \includegraphics[width=0.3\linewidth]{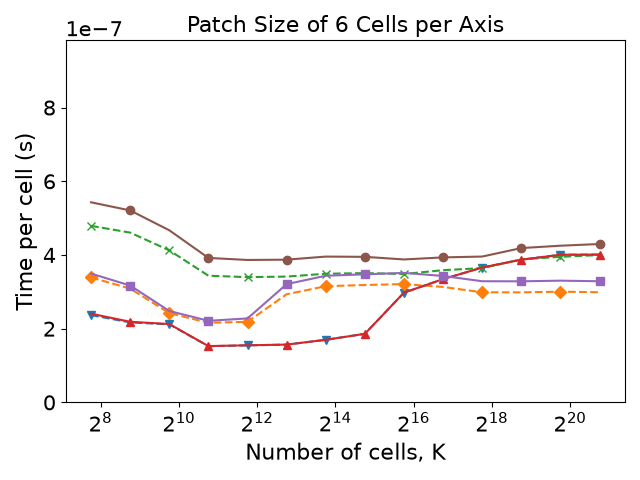}
    \includegraphics[width=0.3\linewidth]{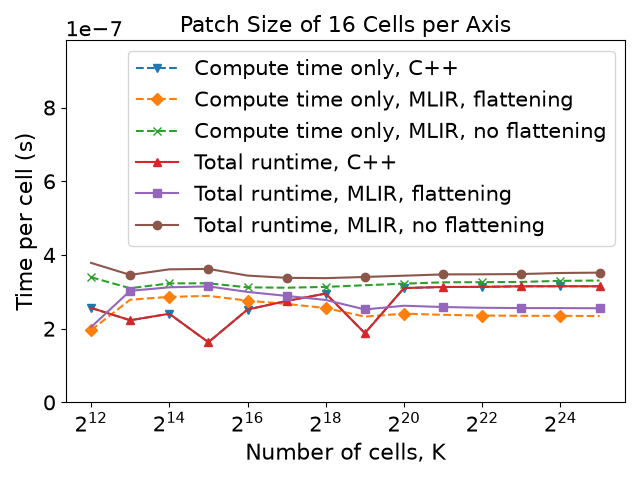}
    \caption{Comparison of the runtime for different serial options for Euler with the FV solver.}
    \label{fig:appendix:plots:euler:rusanov:cpu_serial_comparison:sapphire_rapid}
\end{figure}

\begin{figure}[!htb]
    \centering
    \includegraphics[width=0.3\linewidth]{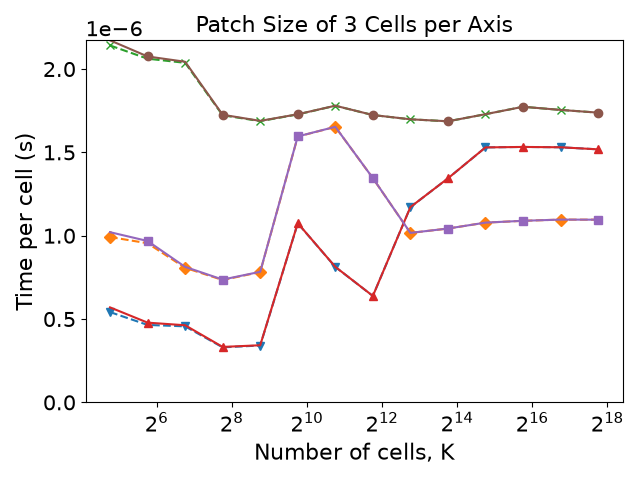}
    \includegraphics[width=0.3\linewidth]{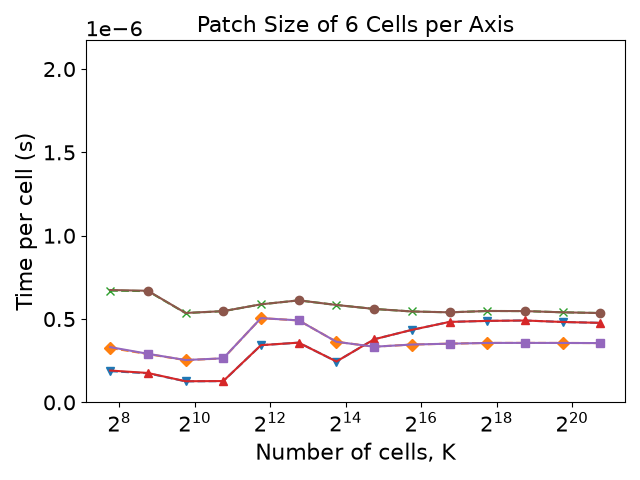}
    \includegraphics[width=0.3\linewidth]{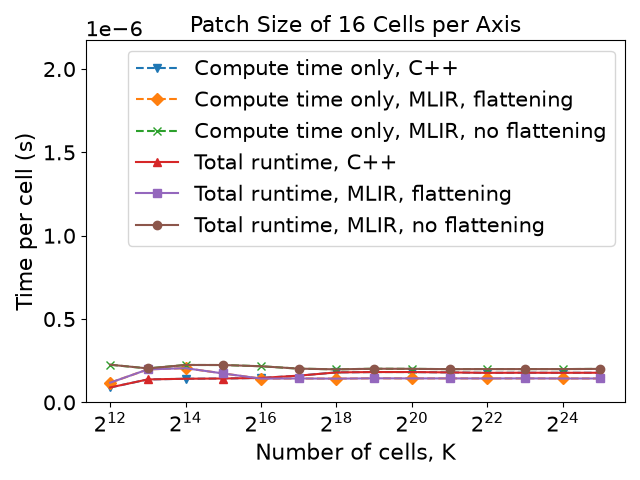}
    \caption{Comparison of the runtime for different serial options for Euler with the FD4 solver.}
    \label{fig:appendix:plots:euler:fd4:cpu_serial_comparison:sapphire_rapid}
\end{figure}

Here we present plots of our serial CPU performance for the Euler equations.
Figures \ref{fig:appendix:plots:euler:rusanov:cpu_serial:sapphire_rapid:normalised} and \ref{fig:appendix:plots:euler:fd4:cpu_serial:sapphire_rapid:normalised} show the differences between the compute time and total runtime per cell for the FV and FD4 solvers respectively.
Figures \ref{fig:appendix:plots:euler:rusanov:cpu_serial:sapphire_rapid:log} and \ref{fig:appendix:plots:euler:fd4:cpu_serial:sapphire_rapid:log} show the same information presented as the sum over all cells. Figures \ref{fig:appendix:plots:euler:rusanov:cpu_serial:sapphire_rapid:all} and \ref{fig:appendix:plots:euler:fd4:cpu_serial:sapphire_rapid:all} show comparisons of the total runtimes across different patch sizes.
Figures \ref{fig:appendix:plots:euler:rusanov:cpu_serial_comparison:sapphire_rapid} and \ref{fig:appendix:plots:euler:fd4:cpu_serial_comparison:sapphire_rapid} show the effect of the MLIR pass that flattens \lstinline!MemRef!s.

\FloatBarrier
\subsubsection{GPU}
\begin{figure}[!htb]
    \centering
    \includegraphics[width=0.3\linewidth]{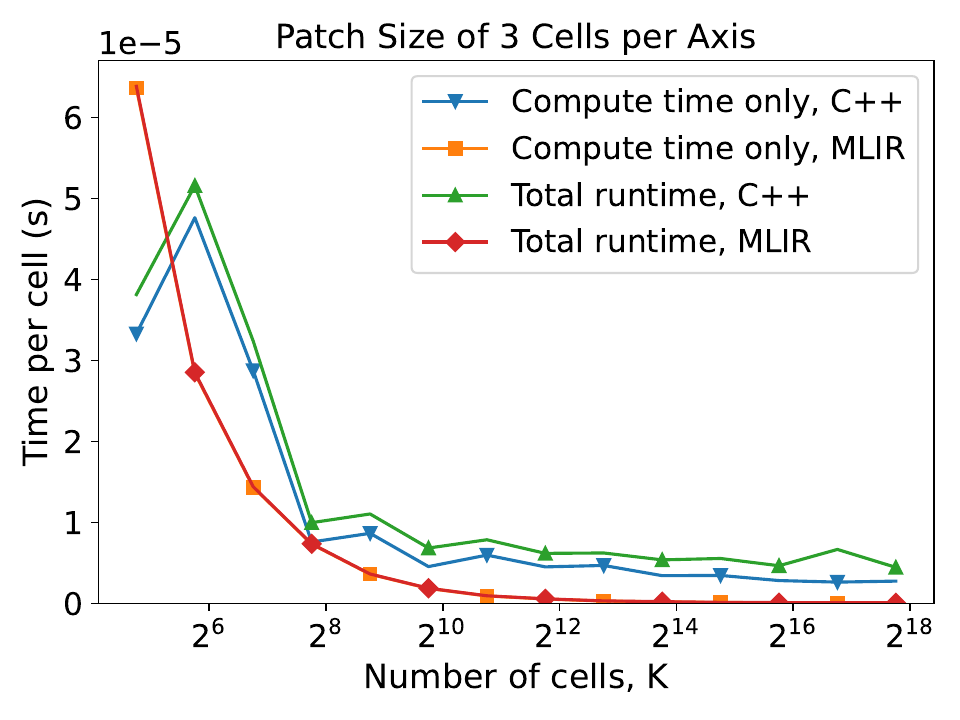}
    \includegraphics[width=0.3\linewidth]{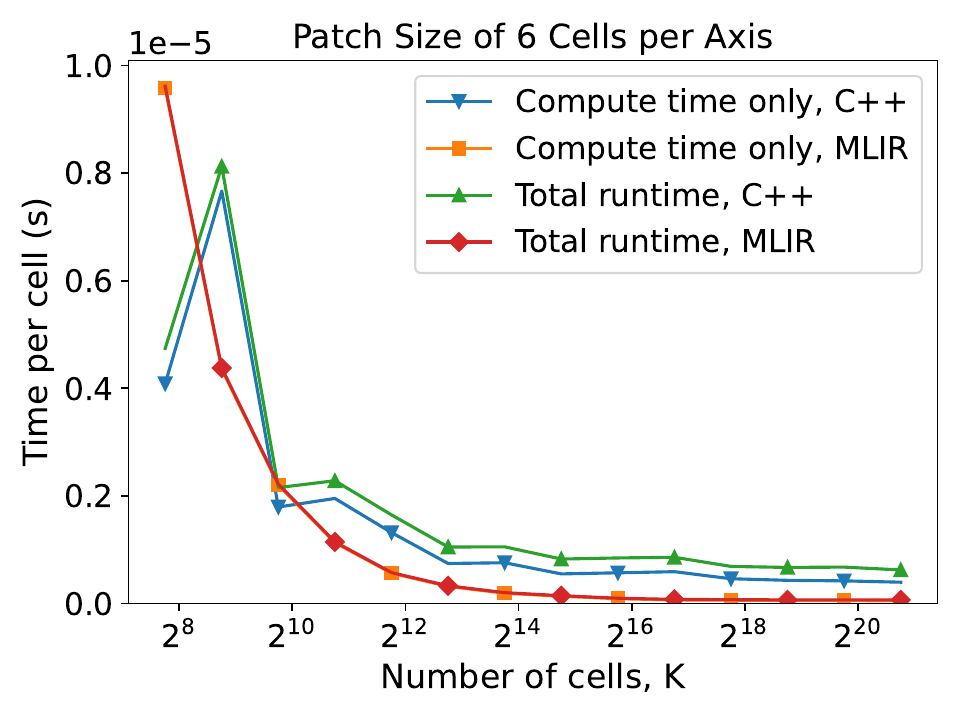}
    \includegraphics[width=0.3\linewidth]{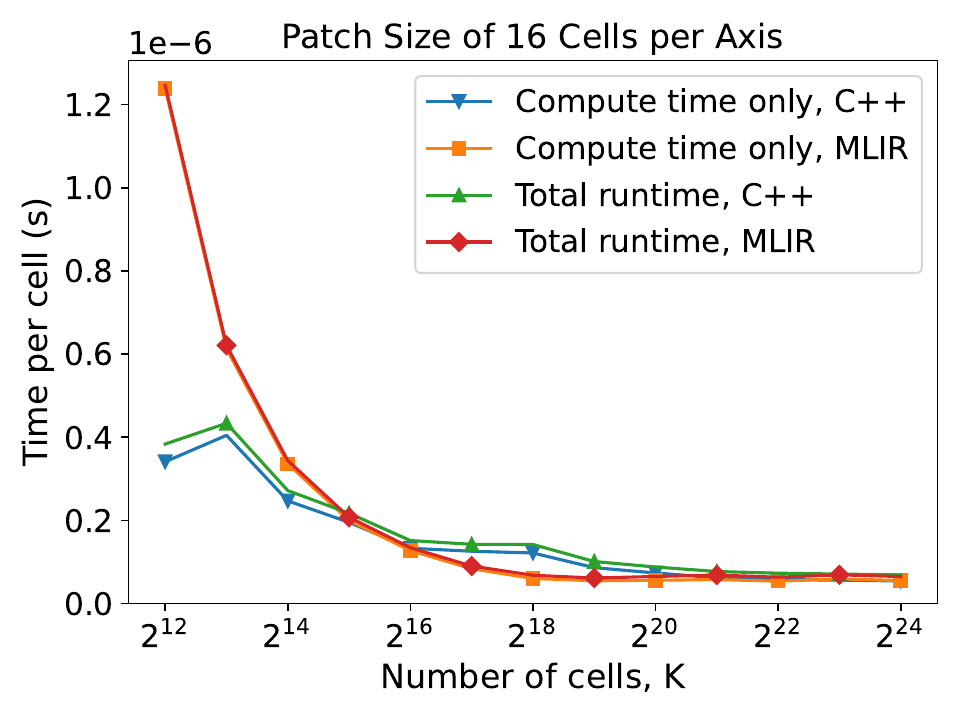}
    \caption{Runtime per cell of Euler with an FV solver}
    \label{fig:appendix:plots:euler:rusanov:gpu_parallel:h200:normalised}
\end{figure}

\begin{figure}[!htb]
    \centering
    \includegraphics[width=0.3\linewidth]{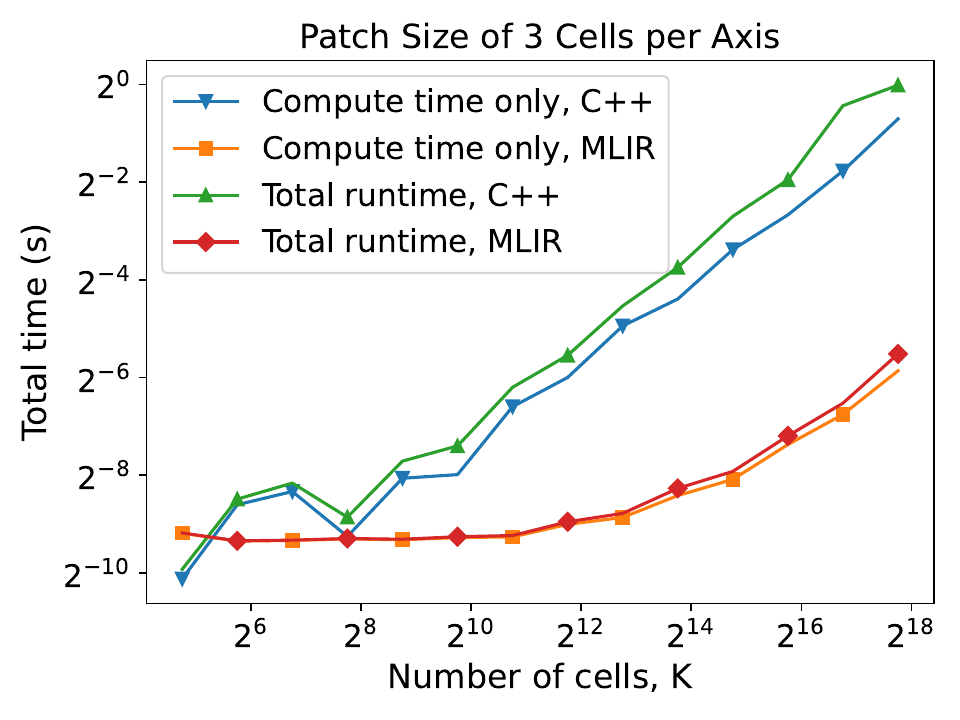}
    \includegraphics[width=0.3\linewidth]{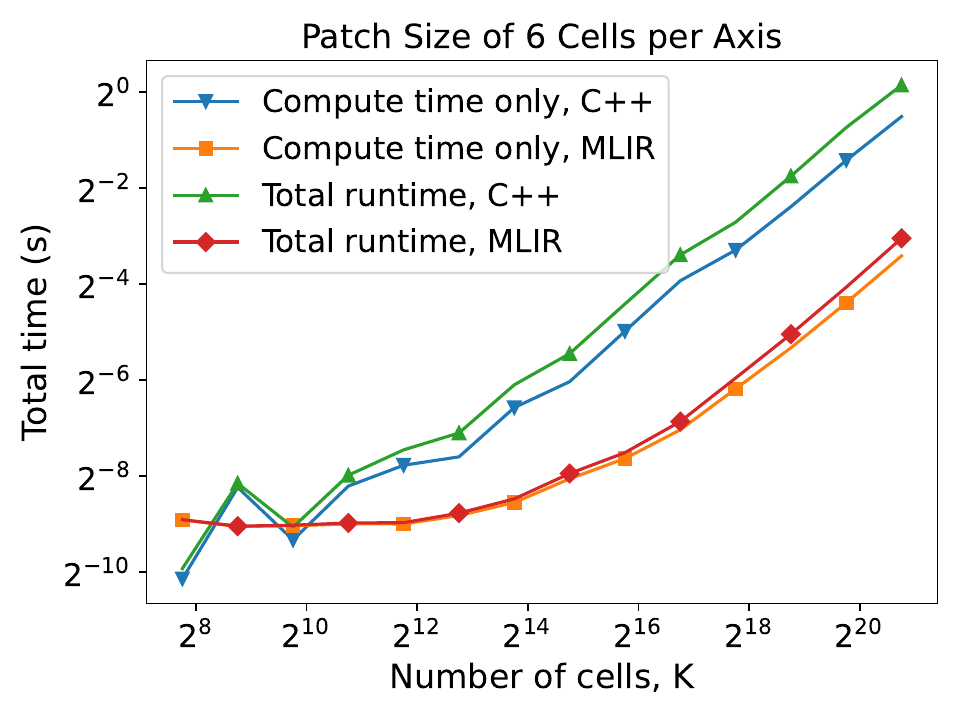}
    \includegraphics[width=0.3\linewidth]{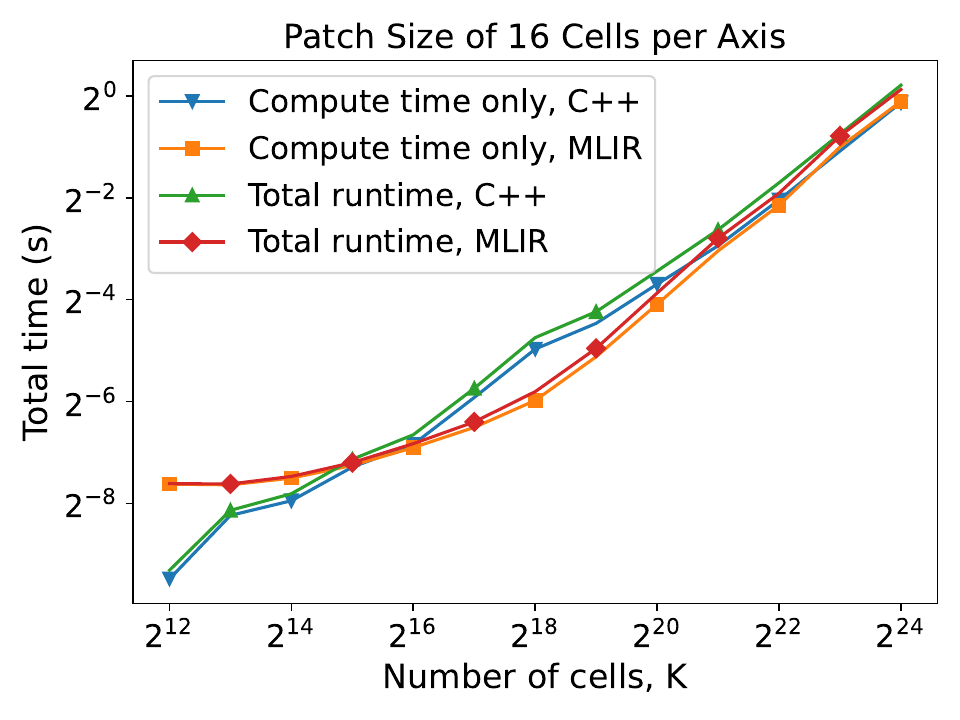}
    \caption{Runtime over all cells of Euler with an FV solver}
    \label{fig:appendix:plots:euler:rusanov:gpu_parallel:h200}
\end{figure}

\begin{figure}[!htb]
    \centering
    \includegraphics[width=0.3\linewidth]{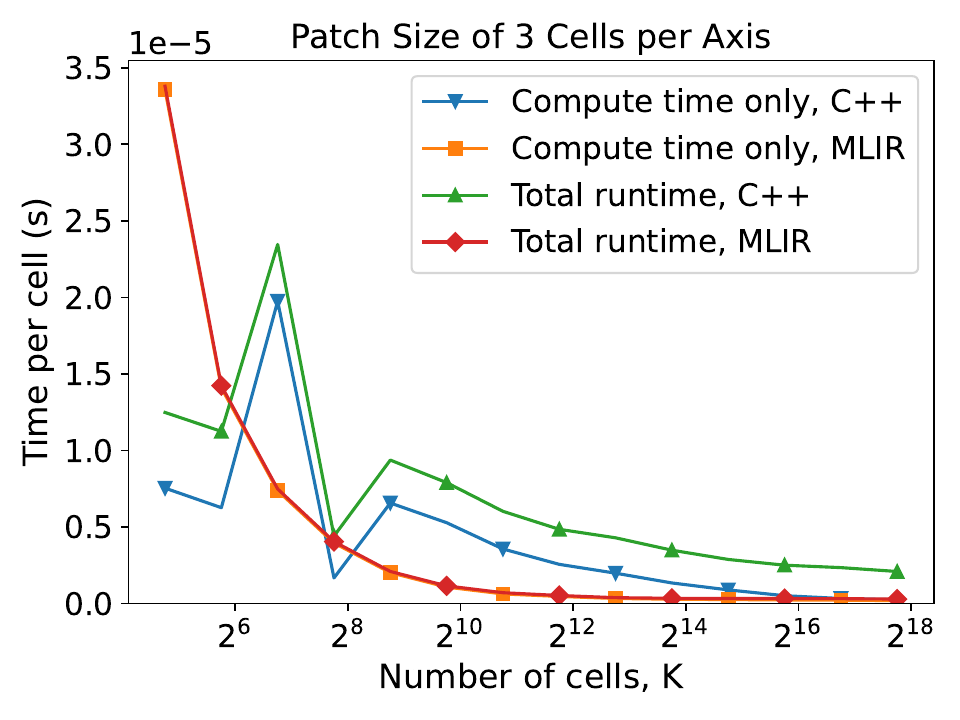}
    \includegraphics[width=0.3\linewidth]{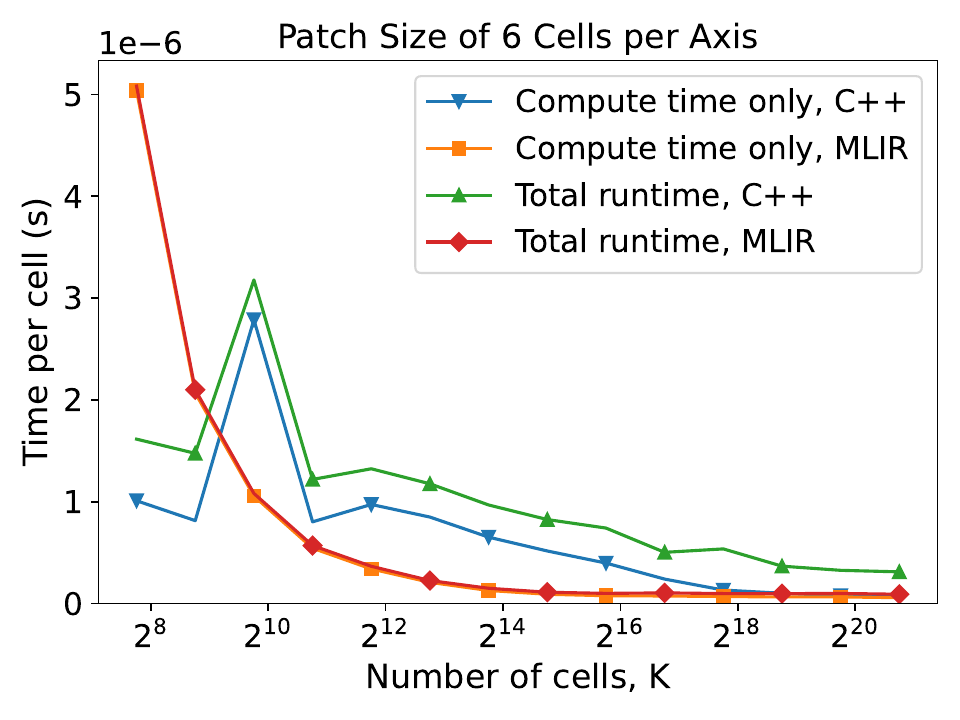}
    \includegraphics[width=0.3\linewidth]{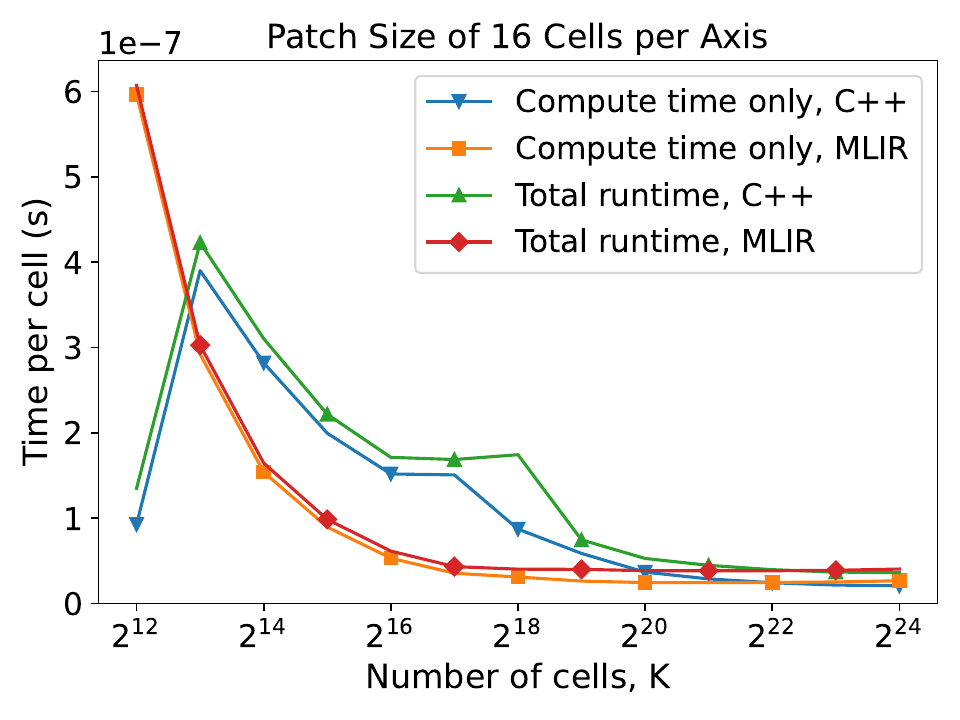}
    \caption{Runtime per cell of Euler with an FD4 solver}
    \label{fig:appendix:plots:euler:fd4:gpu_parallel:h200:normalised}
\end{figure}

\begin{figure}[!htb]
    \centering
    \includegraphics[width=0.3\linewidth]{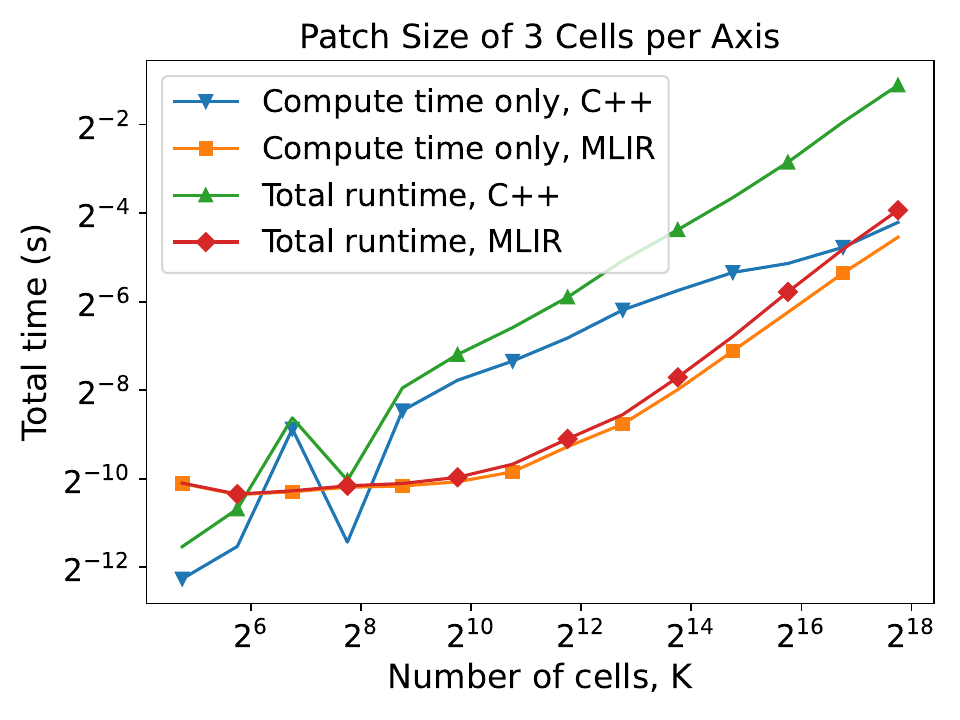}
    \includegraphics[width=0.3\linewidth]{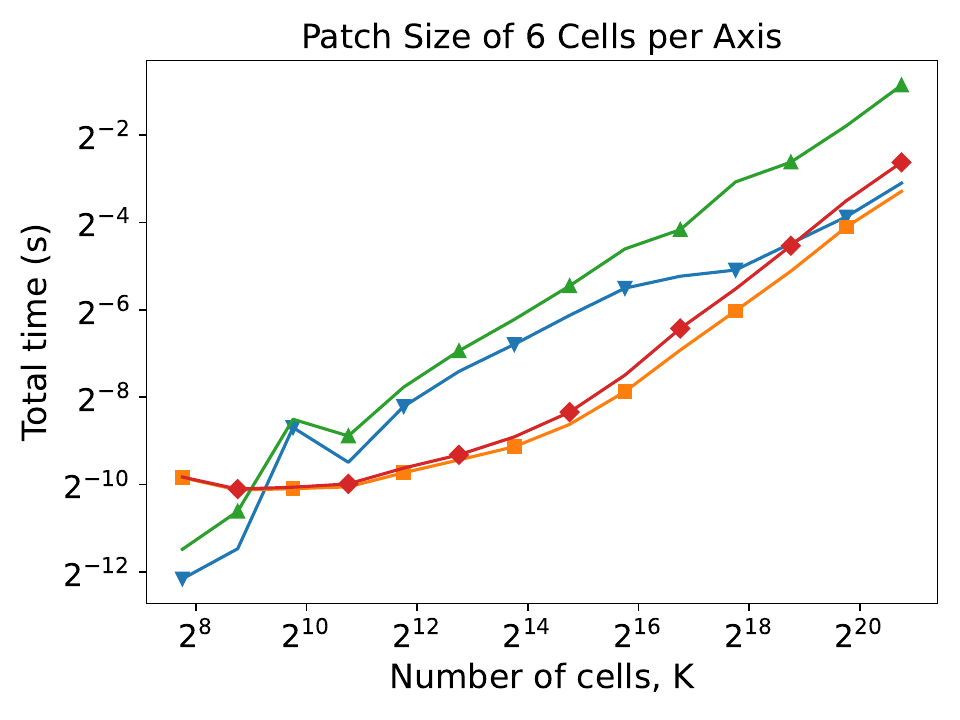}
    \includegraphics[width=0.3\linewidth]{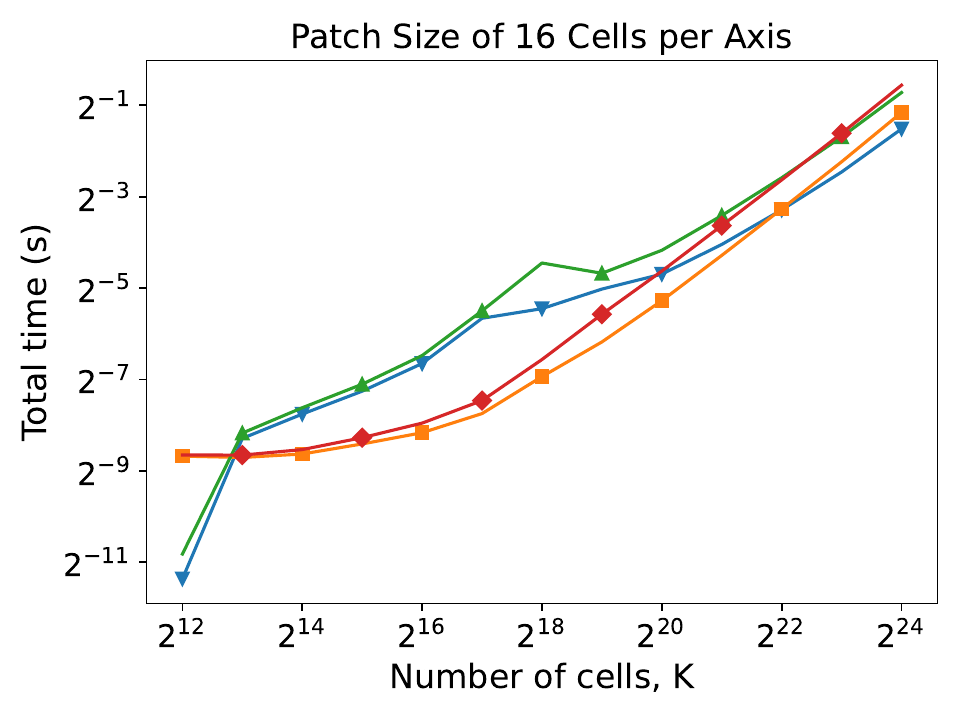}
    \caption{Runtime over all cells of Euler with an FD4 solver}
    \label{fig:appendix:plots:euler:fd4:gpu_parallel:h200}
\end{figure}

Here we present plots of our GPU performance for the Euler equations.
Figures \ref{fig:appendix:plots:euler:rusanov:gpu_parallel:h200:normalised} and \ref{fig:appendix:plots:euler:fd4:gpu_parallel:h200:normalised} show the differences between the compute time and total runtime per cell for the FV and FD4 solvers respectively.
Figures \ref{fig:appendix:plots:euler:rusanov:gpu_parallel:h200} and \ref{fig:appendix:plots:euler:fd4:gpu_parallel:h200} show the same information presented as the sum over all cells.

\FloatBarrier
\subsection{Raw data: CCZ4}
\label{appendix:raw_data_ccz4}
\subsubsection{Serial CPU}
\begin{figure}[!htb]
  \centering
  \includegraphics[width=0.3\linewidth]{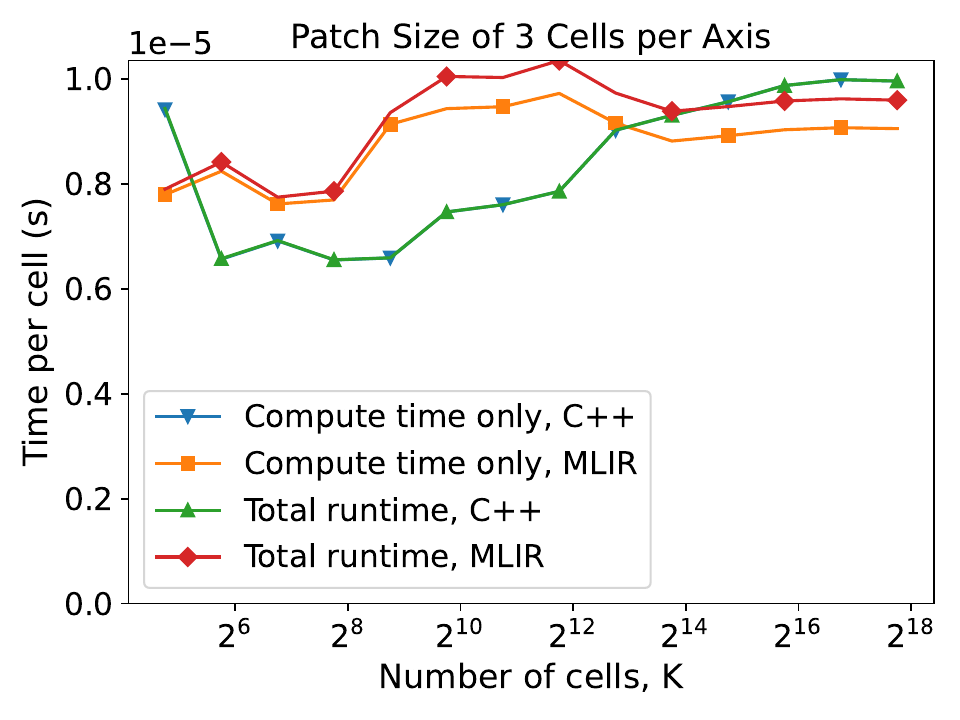}
  \includegraphics[width=0.3\linewidth]{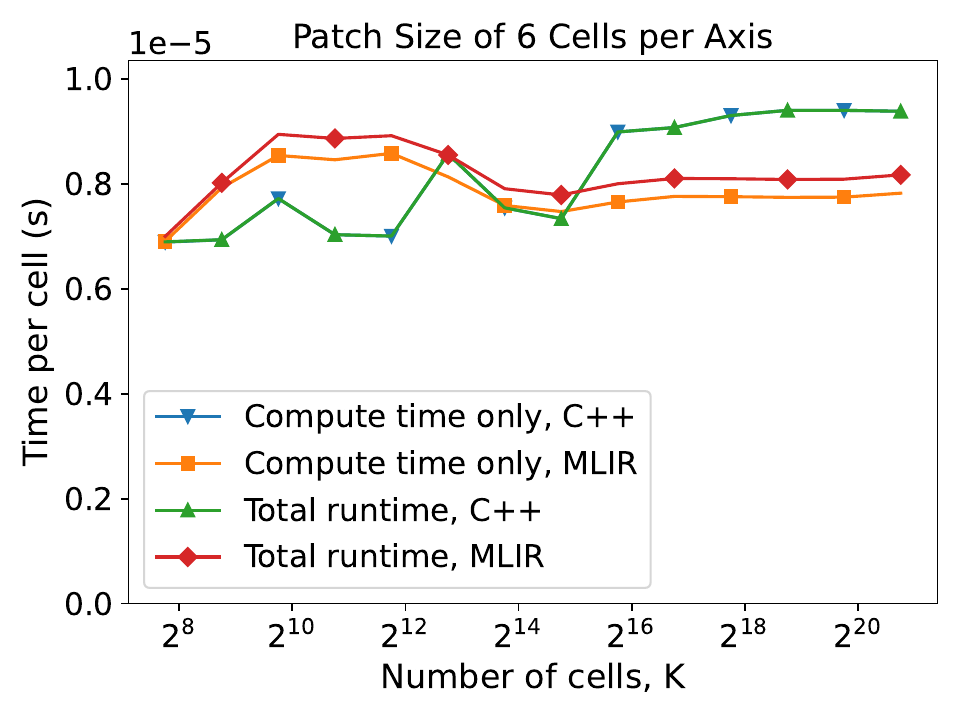}
  \includegraphics[width=0.3\linewidth]{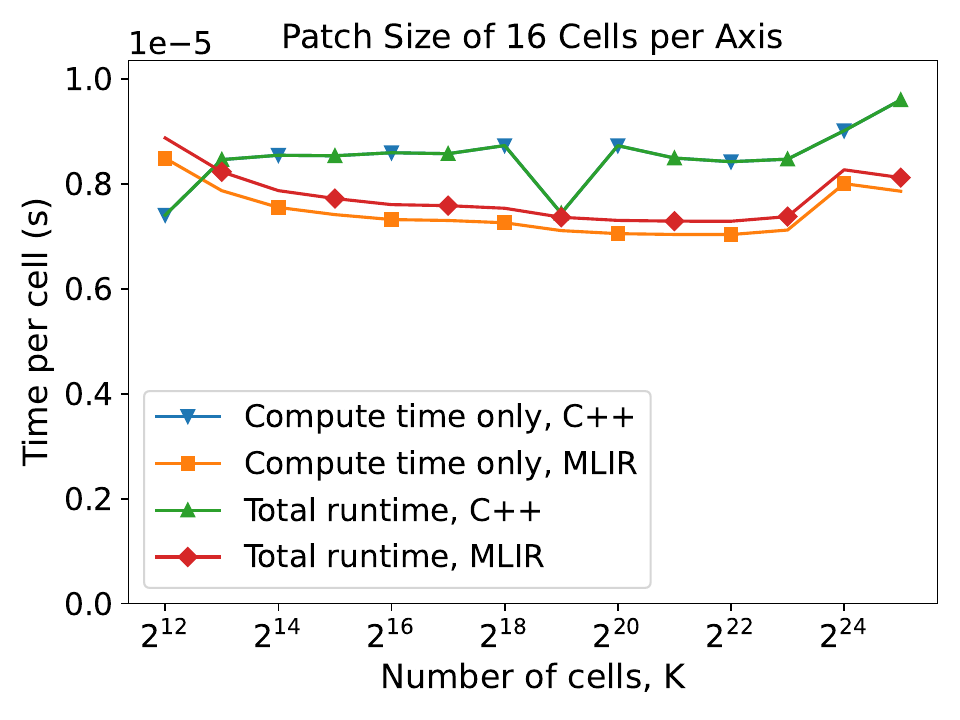}
  \caption{
    Runtime per cell of CCZ4 with Finite Volumes using an FV solver.
}\label{fig:appendix:plots:ccz4:rusanov:cpu_serial:sapphire_rapid:normalised}
\end{figure}

\begin{figure}[!htb]
    \centering
    \includegraphics[width=0.3\linewidth]{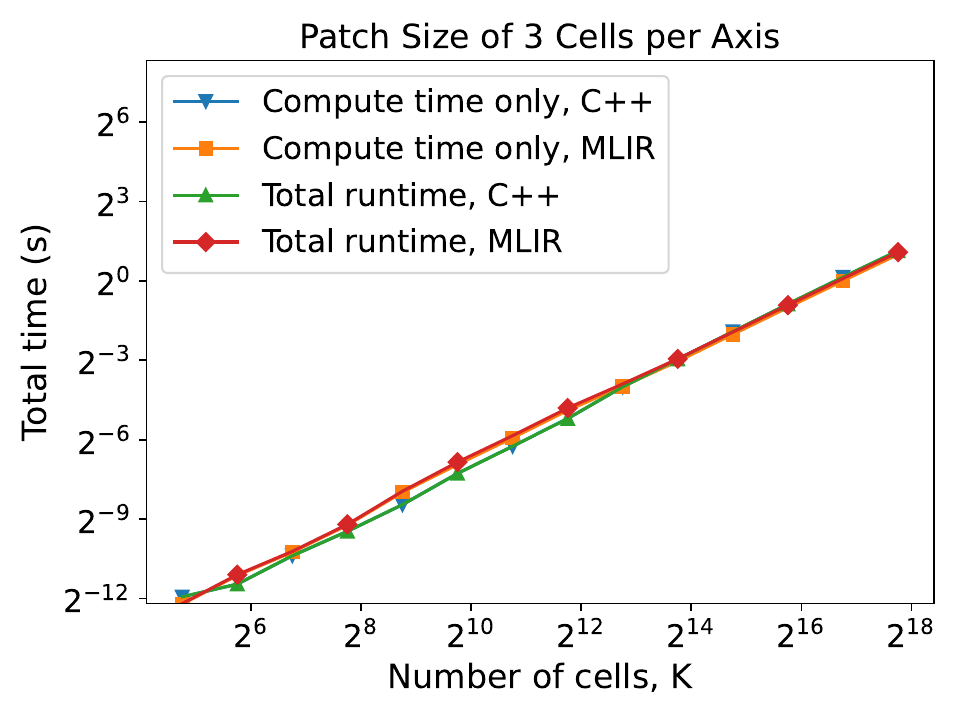}
    \includegraphics[width=0.3\linewidth]{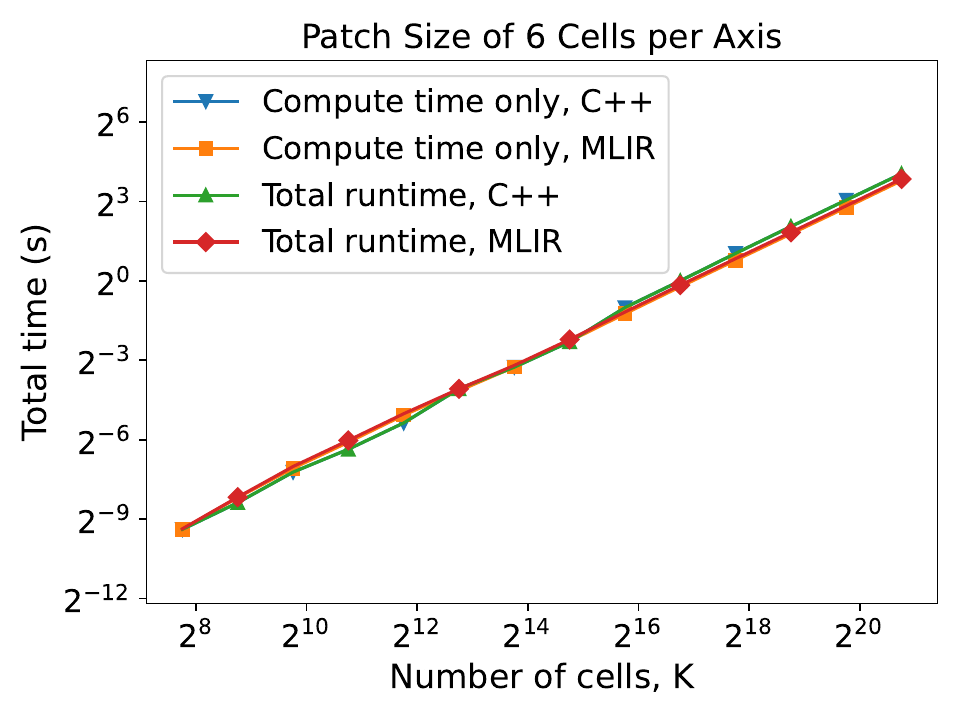}
    \includegraphics[width=0.3\linewidth]{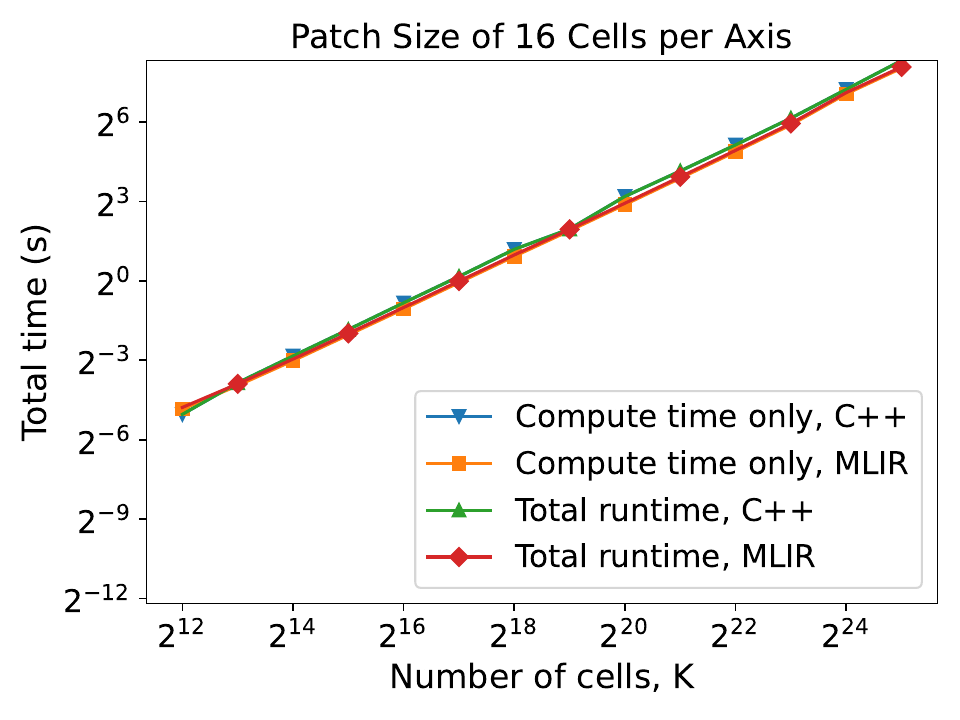}
    \caption{Runtime over all cells of CCZ4 with an FV solver.}
    \label{fig:appendix:plots:ccz4:rusanov:cpu_serial:sapphire_rapid:log}
\end{figure}

\begin{figure}[!htb]
    \centering
    \includegraphics[width=0.4\linewidth]{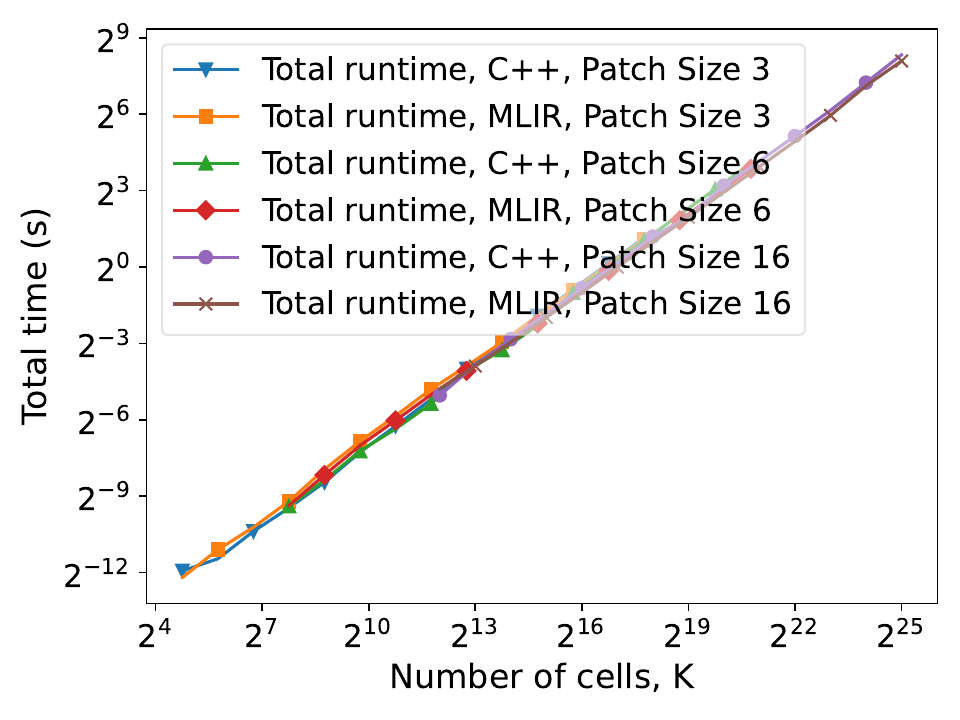}
    \includegraphics[width=0.4\linewidth]{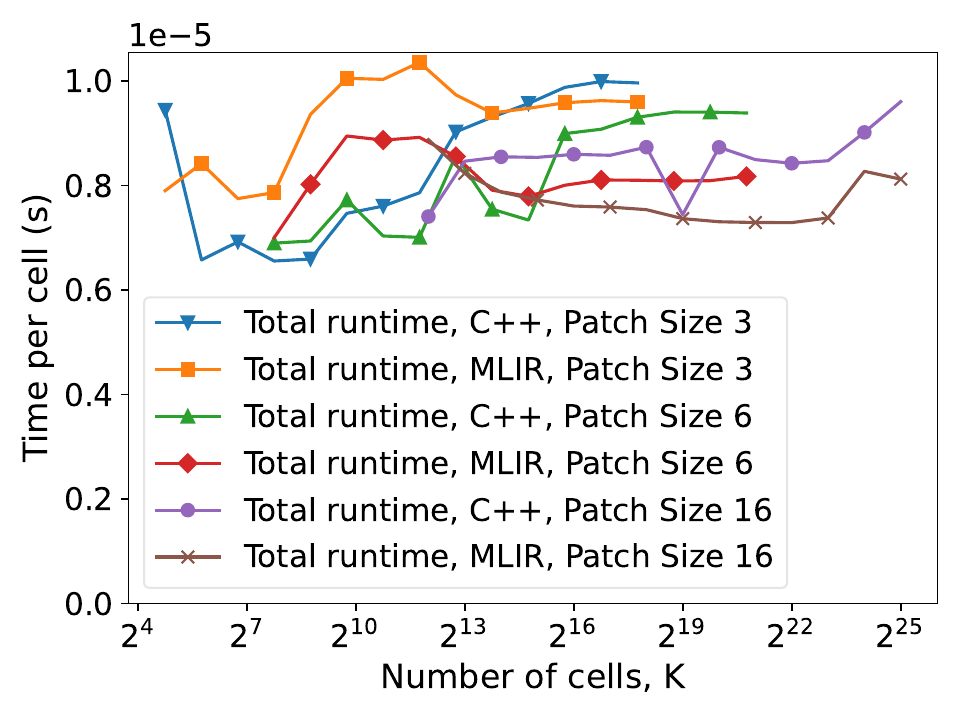}
    \caption{Total runtime for CCZ4 with an FV solver with different patch sizes. Left: Total runtime over all cells, right: runtime per cell.}
    \label{fig:appendix:plots:ccz4:rusanov:cpu_serial:sapphire_rapid:all}
\end{figure}

\begin{figure}[!htb]
  \centering
  \includegraphics[width=0.3\linewidth]{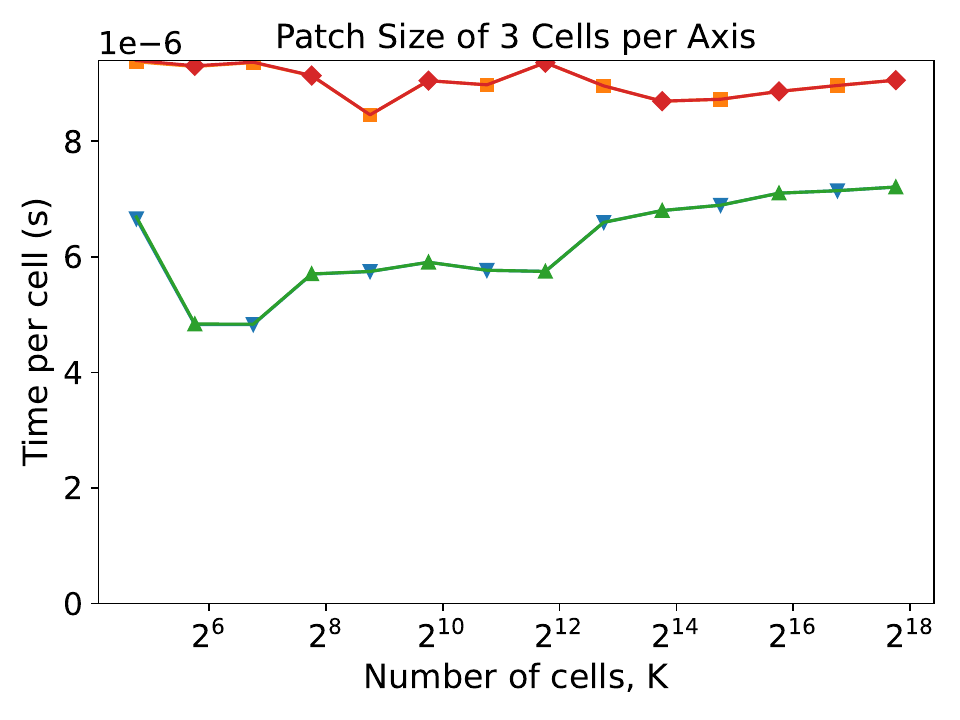}
  \includegraphics[width=0.3\linewidth]{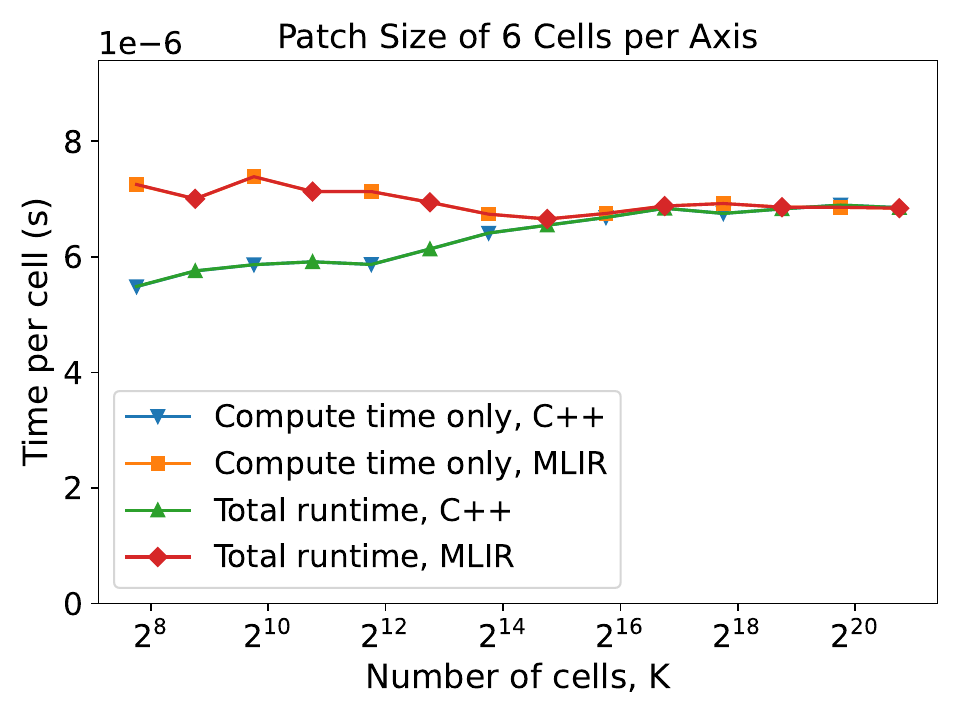}
  \includegraphics[width=0.3\linewidth]{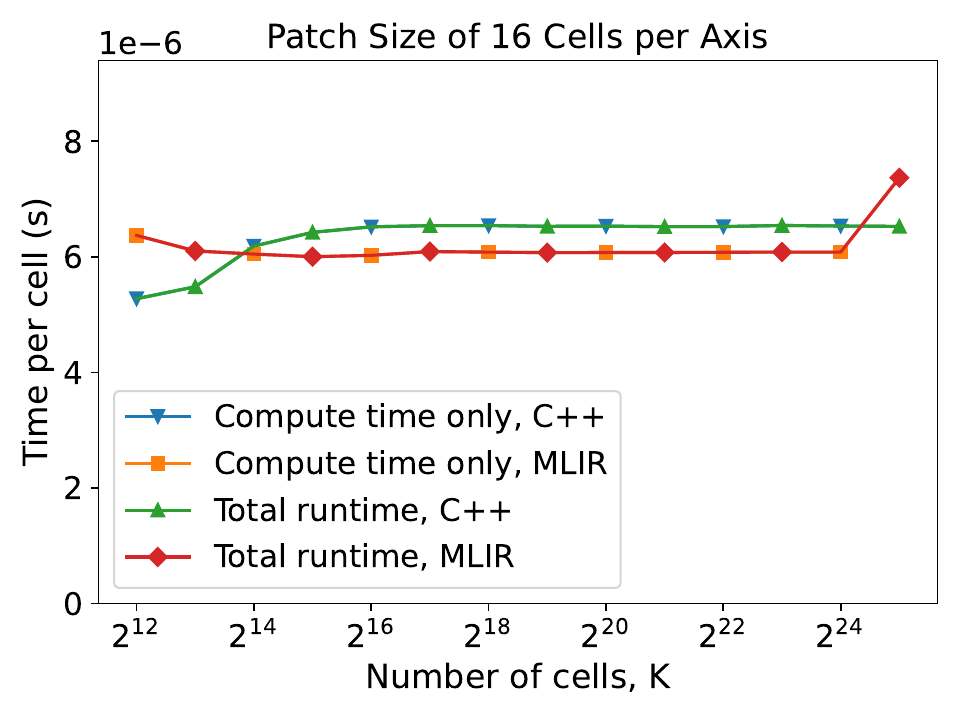}
  \caption{
    Runtime per cell of CCZ4 with Finite Volumes using an FD4 solver. \label{fig:appendix:plots:ccz4:fd4:cpu_serial:sapphire_rapid:normalised}
  }
\end{figure}

\begin{figure}[!htb]
    \centering
    \includegraphics[width=0.3\linewidth]{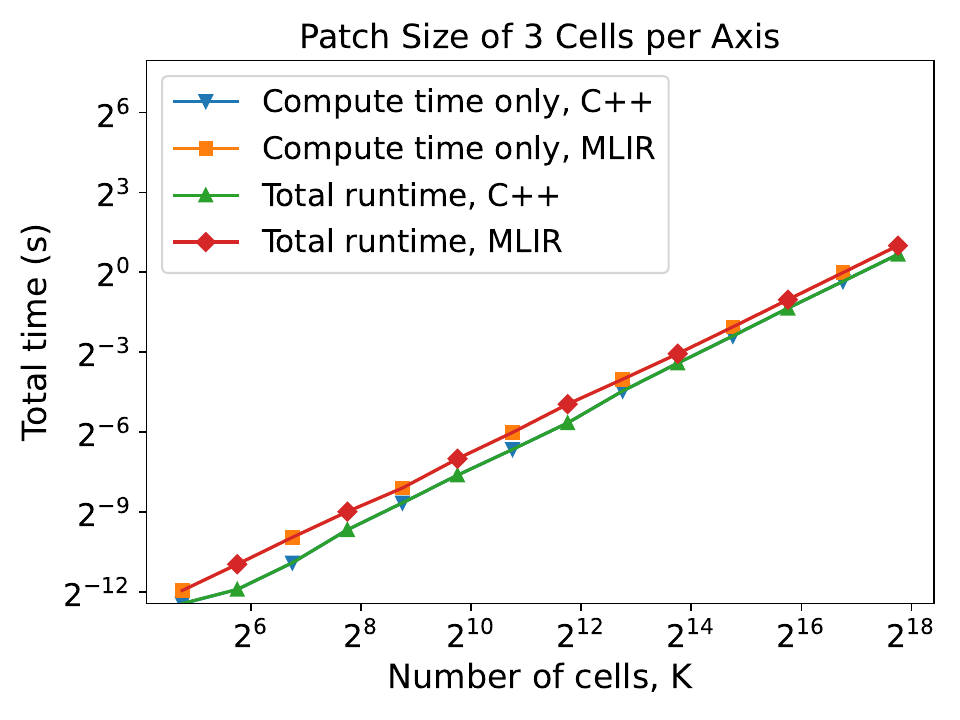}
    \includegraphics[width=0.3\linewidth]{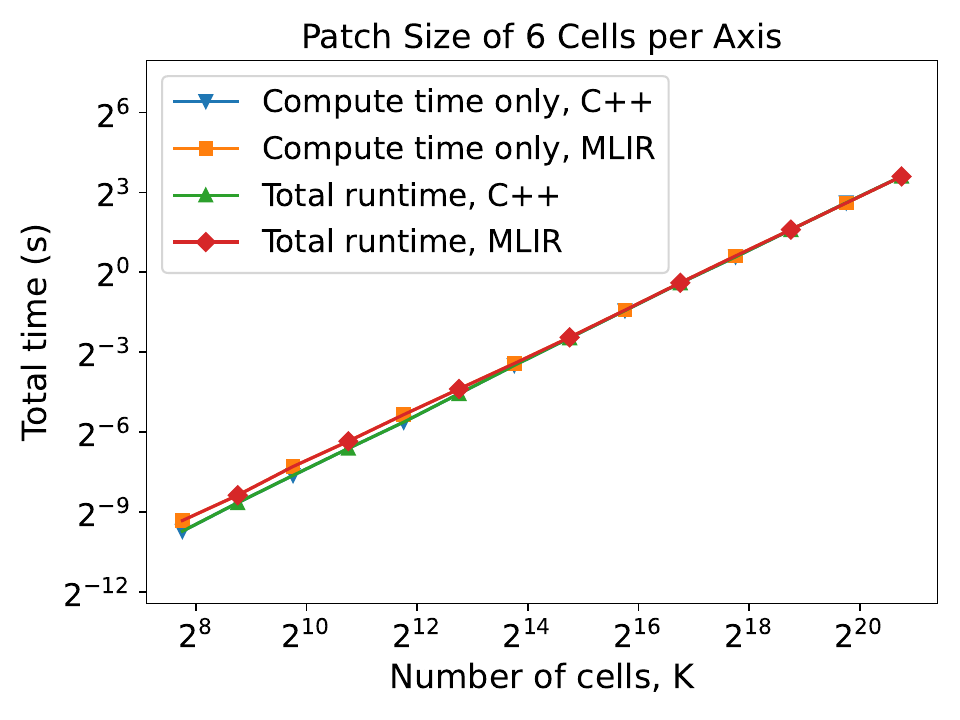}
    \includegraphics[width=0.3\linewidth]{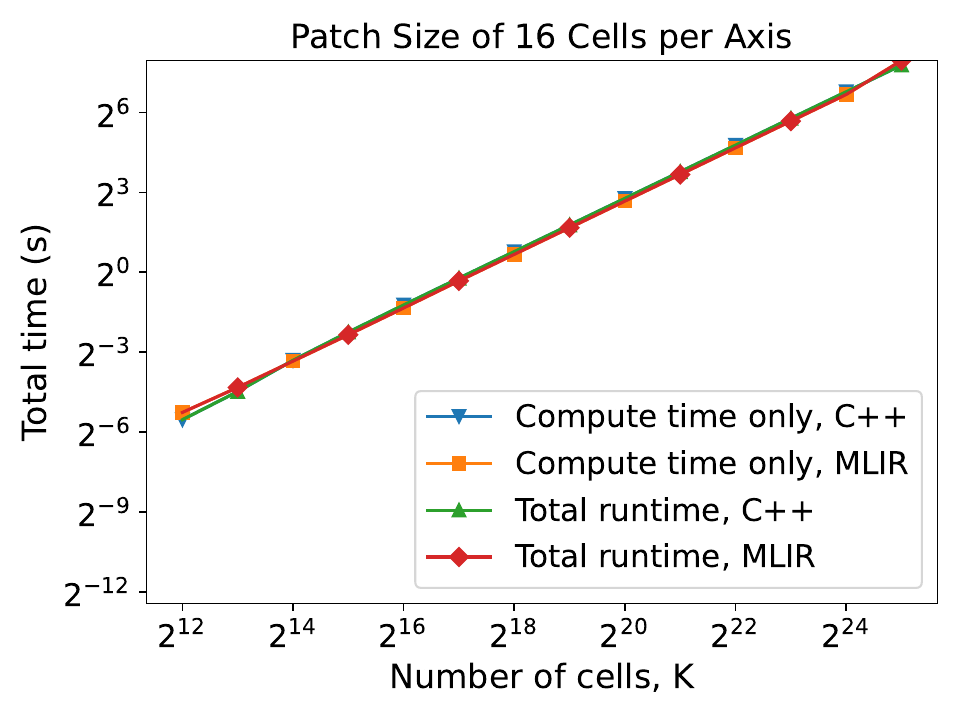}
    \caption{Runtime over all cells of CCZ4 with Finite Volumes using an FD4 solver.}
    \label{fig:appendix:plots:ccz4:fd4:cpu_serial:sapphire_rapid:log}
\end{figure}

\begin{figure}[!htb]
    \centering
    \includegraphics[width=0.4\linewidth]{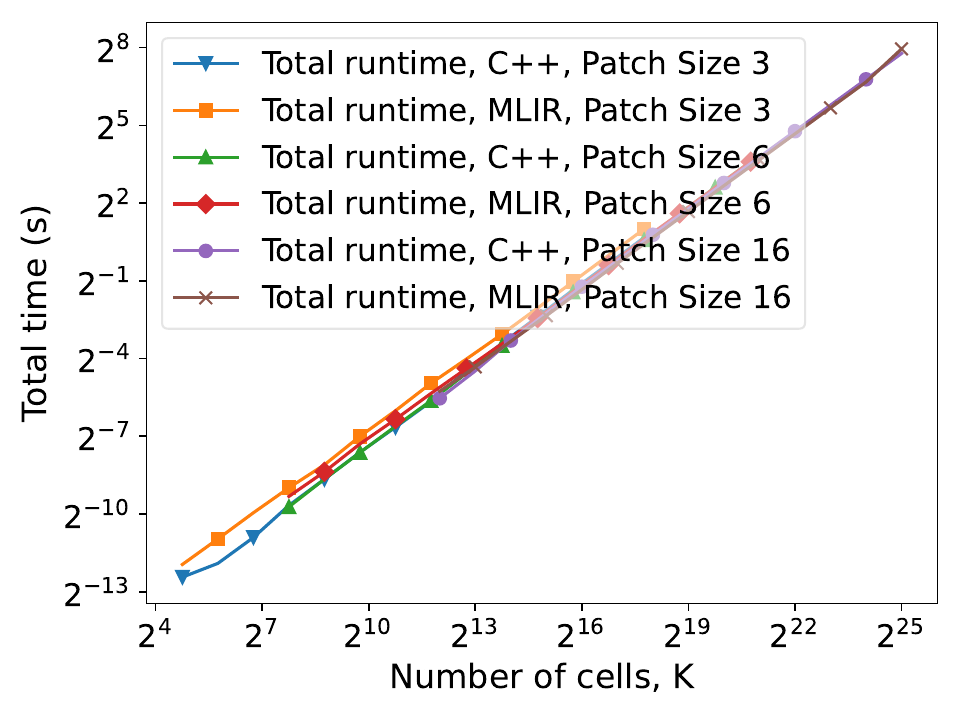}
    \includegraphics[width=0.4\linewidth]{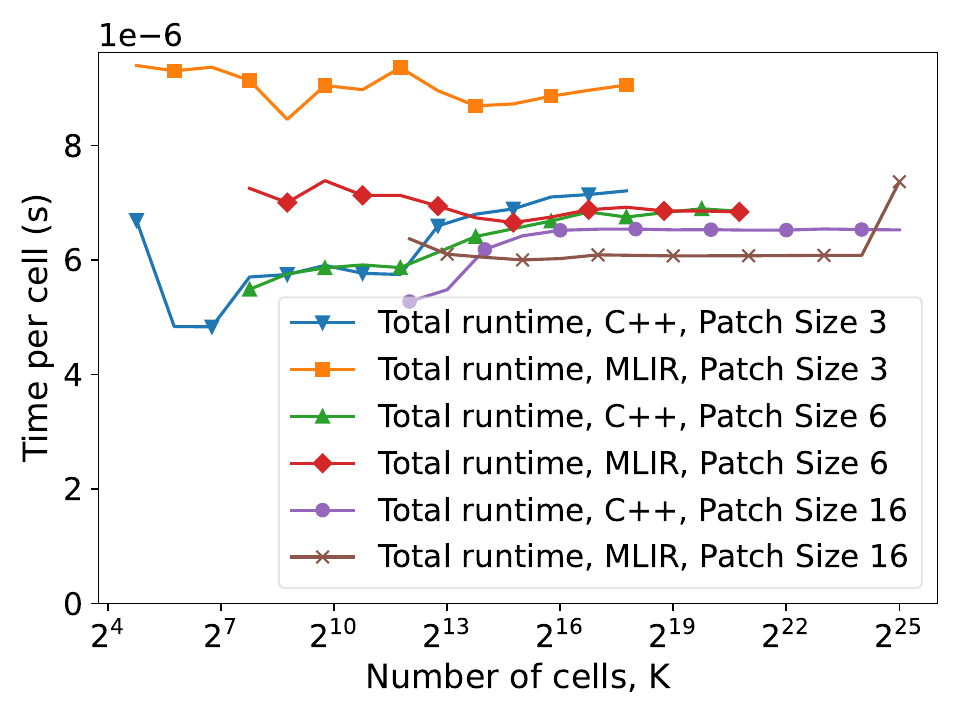}
    \caption{Total runtime for CCZ4 with an FD4 solver with different patch sizes. Left: Total runtime over all cells, right: runtime per cell.}
    \label{fig:appendix:plots:ccz4:fd4:cpu_serial:sapphire_rapid:all}
\end{figure}

\begin{figure}[!htb]
    \centering
    \includegraphics[width=0.3\linewidth]{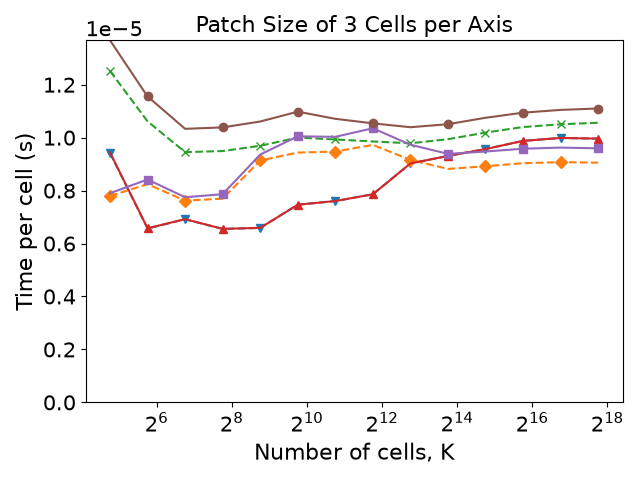}
    \includegraphics[width=0.3\linewidth]{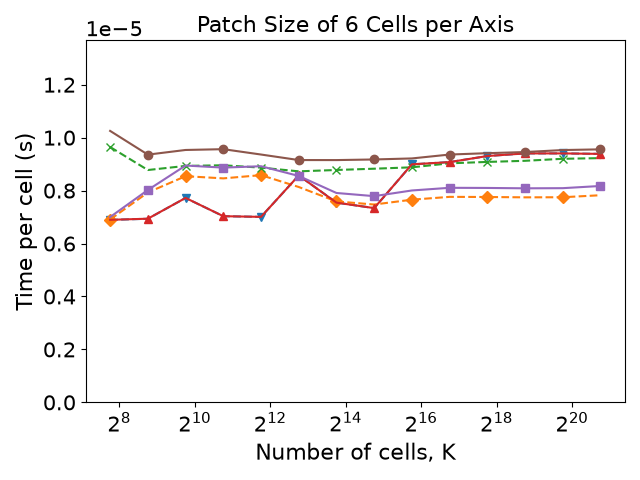}
    \includegraphics[width=0.3\linewidth]{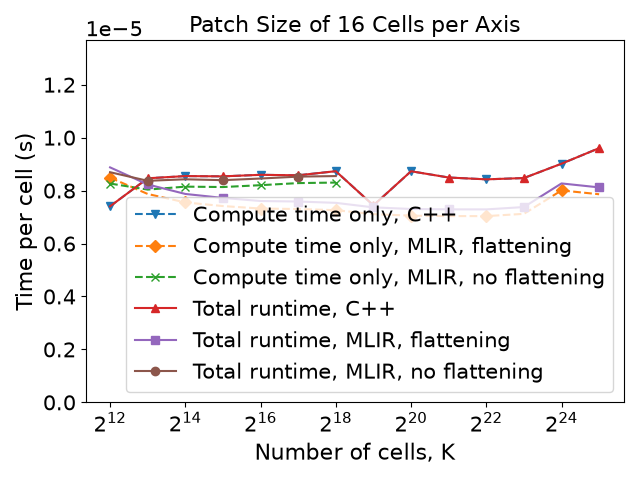}
    \caption{Comparison of the runtime for different serial options for CCZ4 with the FV solver.}
    \label{fig:appendix:plots:ccz4:rusanov:cpu_serial_comparison:sapphire_rapid}
\end{figure}

\begin{figure}[!htb]
    \centering
    \includegraphics[width=0.3\linewidth]{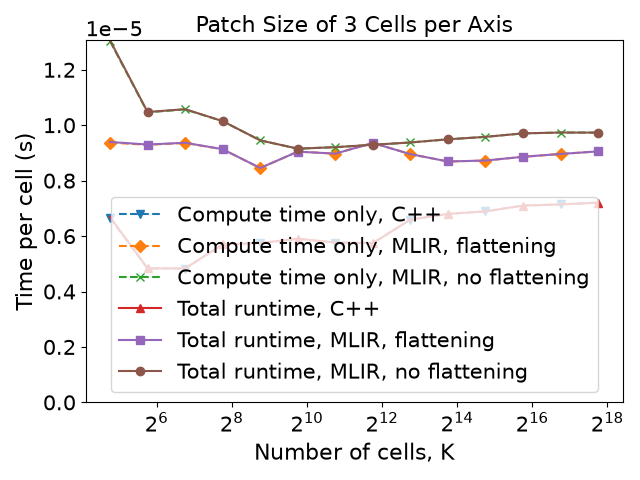}
    \includegraphics[width=0.3\linewidth]{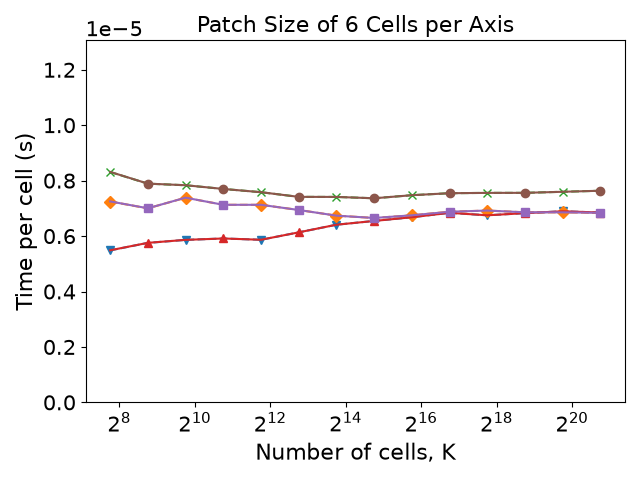}
    \includegraphics[width=0.3\linewidth]{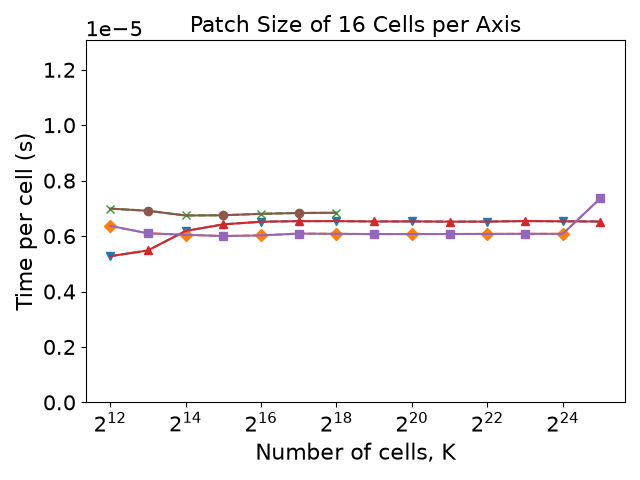}
    \caption{Comparison of the runtime for different serial options for CCZ4 with the FD4 solver.}
    \label{fig:appendix:plots:ccz4:fd4:cpu_serial_comparison:sapphire_rapid}
\end{figure}

Here we present plots of our serial CPU performance for the CCZ4 equations.
Figures \ref{fig:appendix:plots:ccz4:rusanov:cpu_serial:sapphire_rapid:normalised} and \ref{fig:appendix:plots:ccz4:fd4:cpu_serial:sapphire_rapid:normalised} show the differences between the compute time and total runtime per cell for the FV and FD4 solvers respectively.
Figures \ref{fig:appendix:plots:ccz4:rusanov:cpu_serial:sapphire_rapid:log} and \ref{fig:appendix:plots:ccz4:fd4:cpu_serial:sapphire_rapid:log} show the same information presented as the sum over all cells. Figures \ref{fig:appendix:plots:ccz4:rusanov:cpu_serial:sapphire_rapid:all} and \ref{fig:appendix:plots:ccz4:fd4:cpu_serial:sapphire_rapid:all} show comparisons of the total runtimes across different patch sizes.
Figures \ref{fig:appendix:plots:ccz4:rusanov:cpu_serial_comparison:sapphire_rapid} and \ref{fig:appendix:plots:ccz4:fd4:cpu_serial_comparison:sapphire_rapid} show the effect of the MLIR pass that flattens \lstinline!MemRef!s.

\FloatBarrier
\subsubsection{GPU}
\begin{figure}[!htb]
    \centering
    \includegraphics[width=0.3\linewidth]{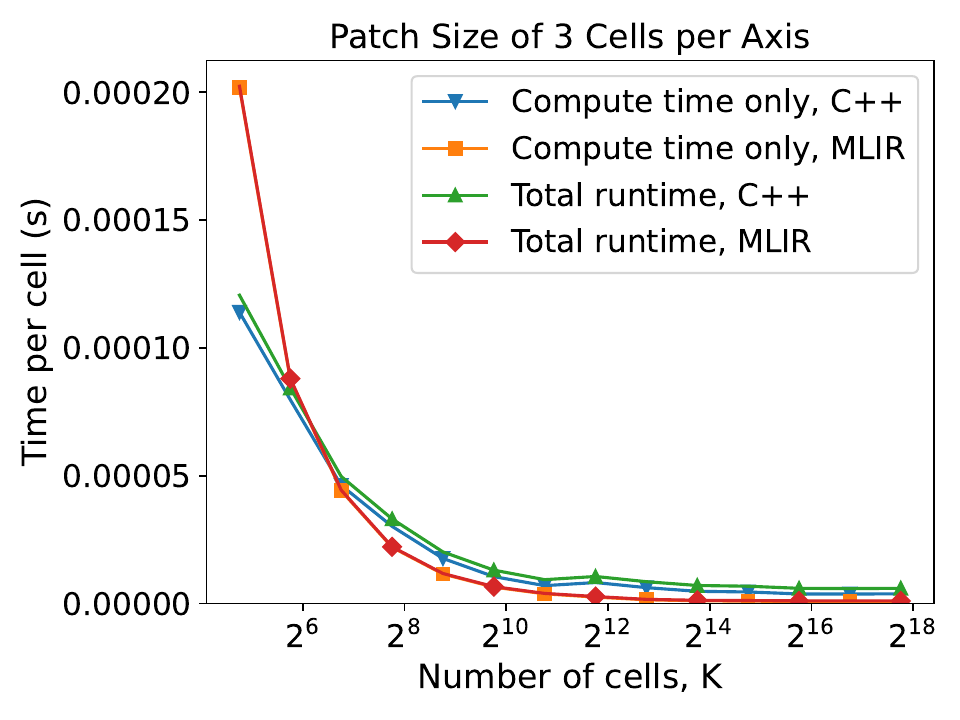}
    \includegraphics[width=0.3\linewidth]{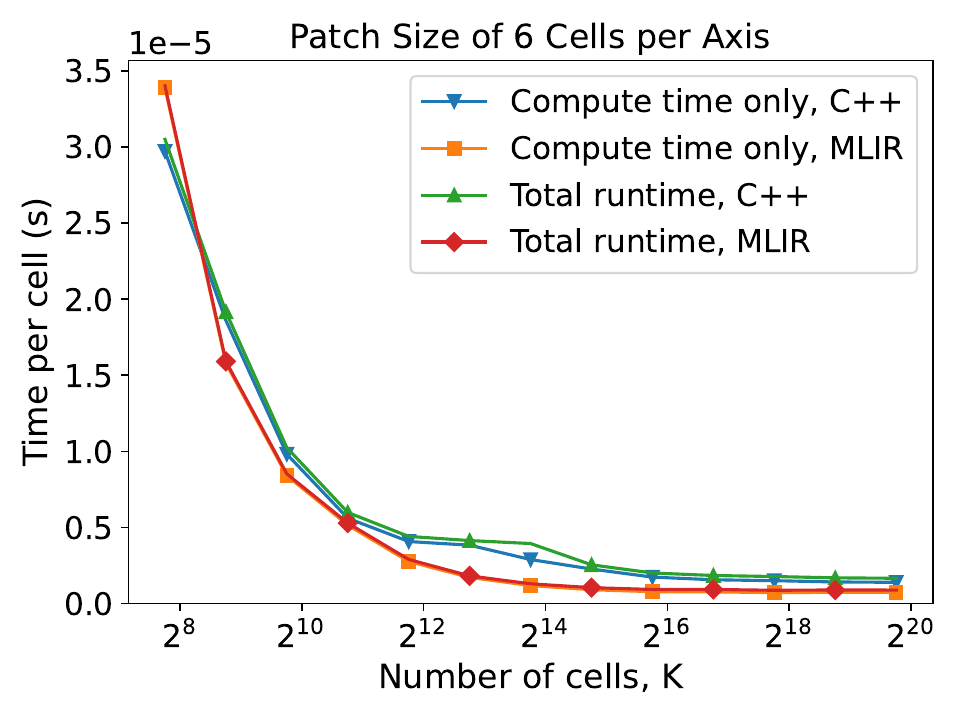}
    \includegraphics[width=0.3\linewidth]{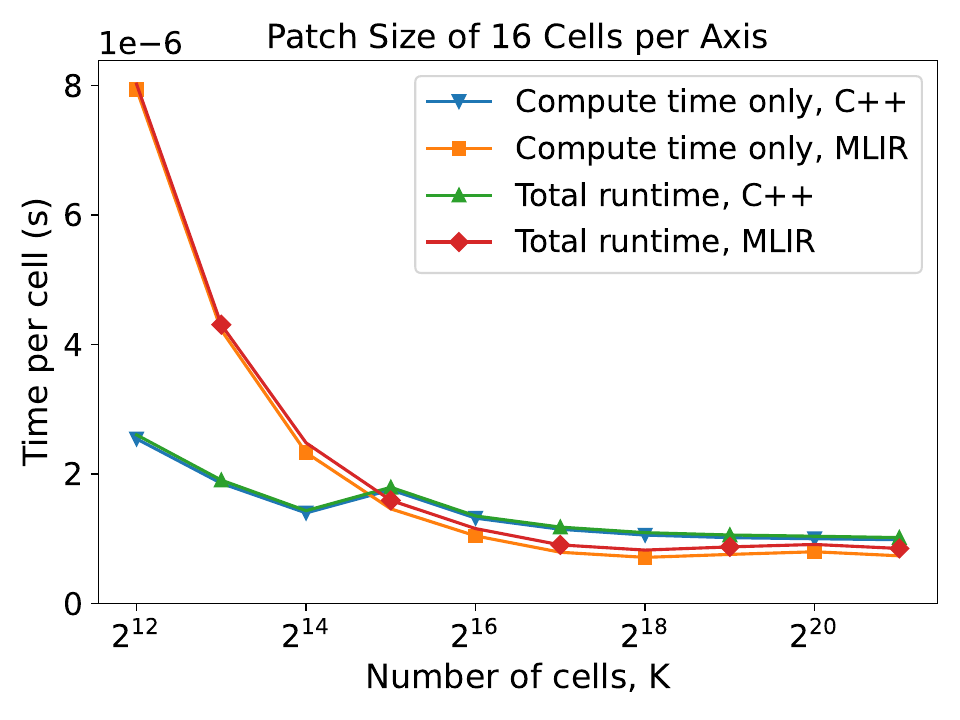}
    \caption{Runtime per cell of CCZ4 with an FV solver}
    \label{fig:appendix:plots:ccz4:rusanov:gpu_parallel:h200:normalised}
\end{figure}

\begin{figure}[!htb]
    \centering
    \includegraphics[width=0.3\linewidth]{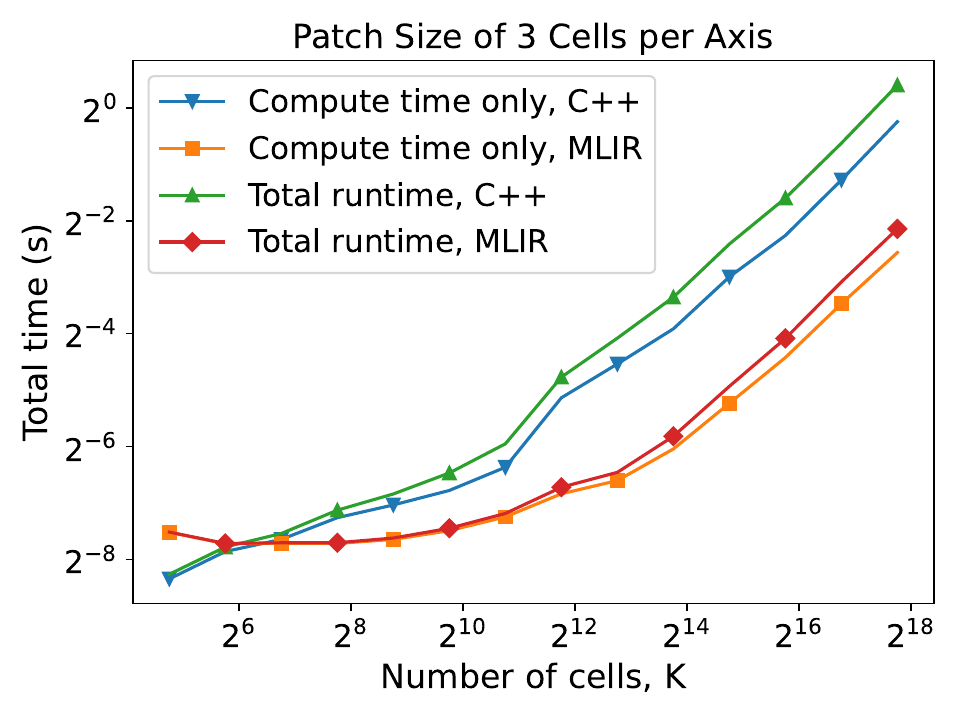}
    \includegraphics[width=0.3\linewidth]{plots/ccz4/rusanov/gpu_parallel/flattened_memrefs_on/grace_hopper/patch_size_6_with_legend.pdf}
    \includegraphics[width=0.3\linewidth]{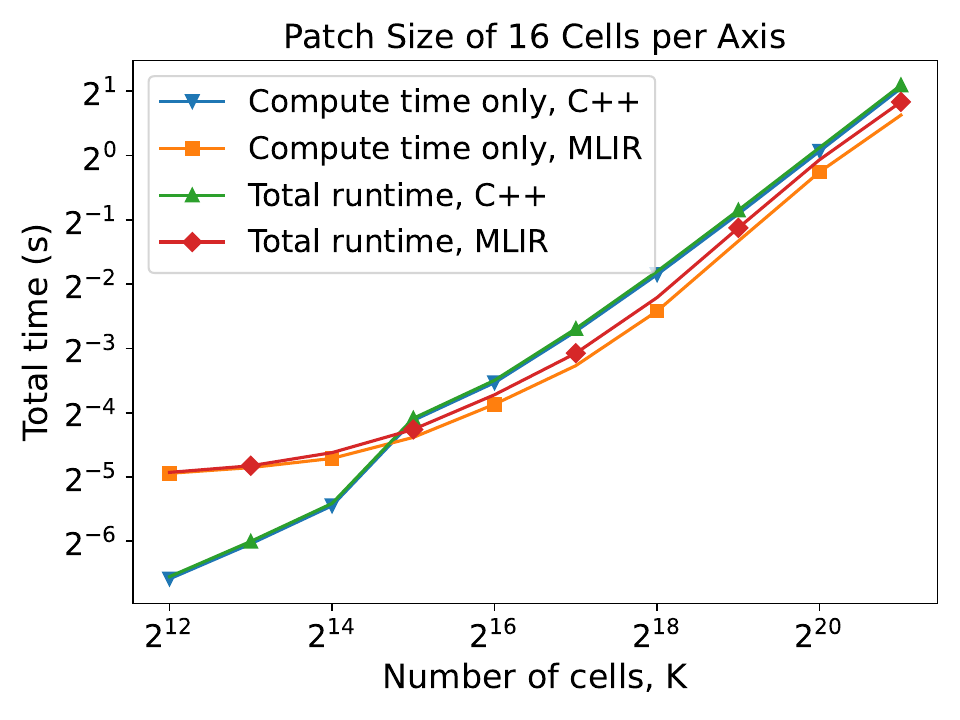}
    \caption{Runtime over all cells of CCZ4 with an FV solver}
    \label{fig:appendix:plots:ccz4:rusanov:gpu_parallel:h200}
\end{figure}

\begin{figure}[!htb]
    \centering
    \includegraphics[width=0.3\linewidth]{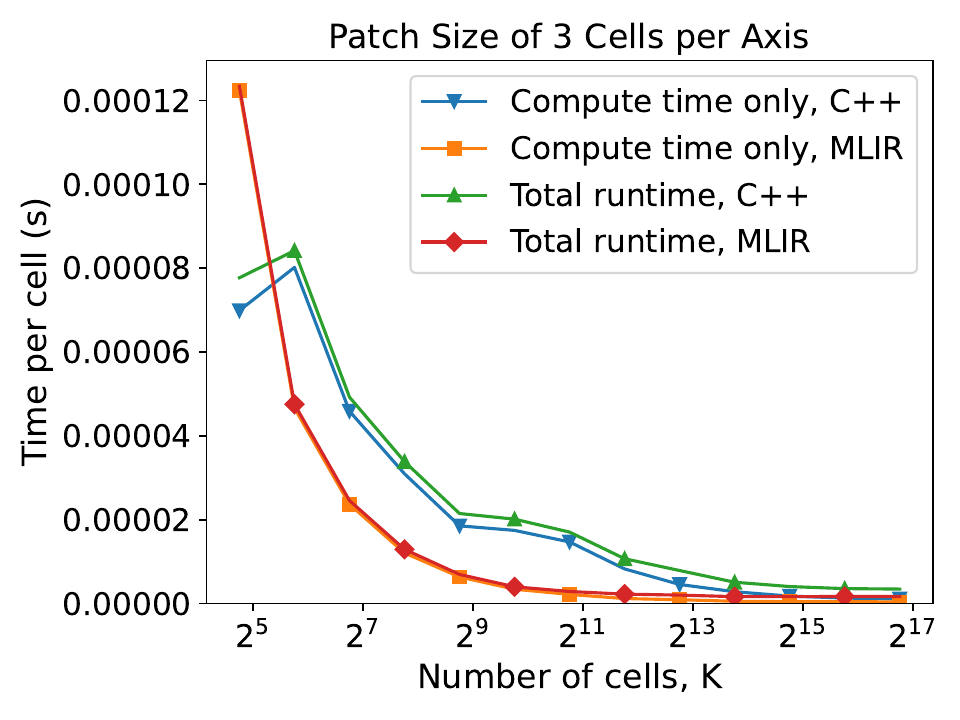}
    \includegraphics[width=0.3\linewidth]{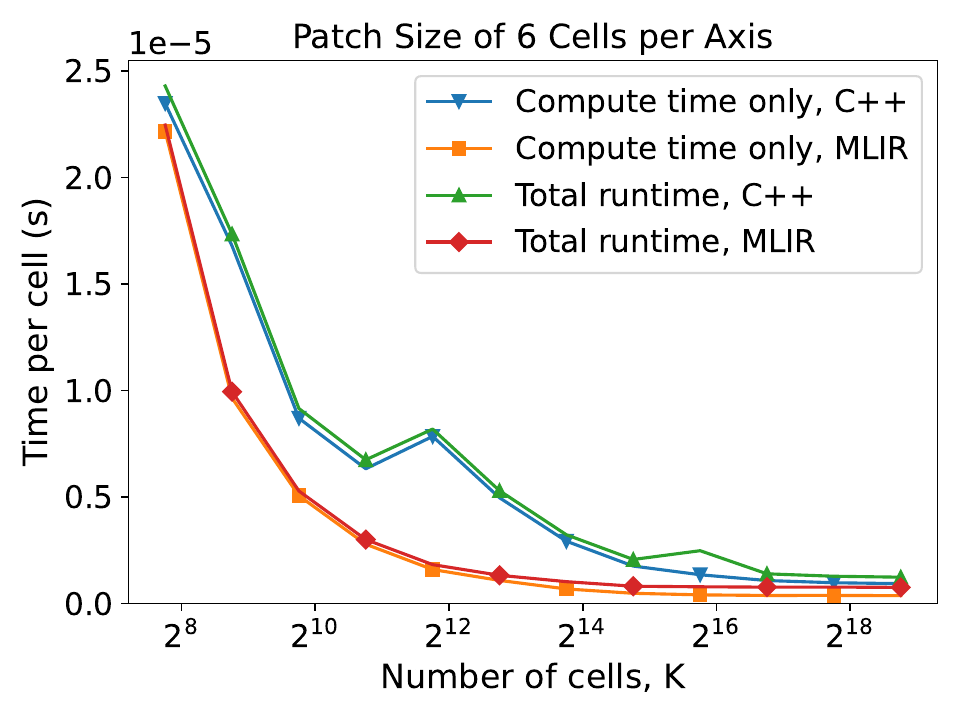}
    \includegraphics[width=0.3\linewidth]{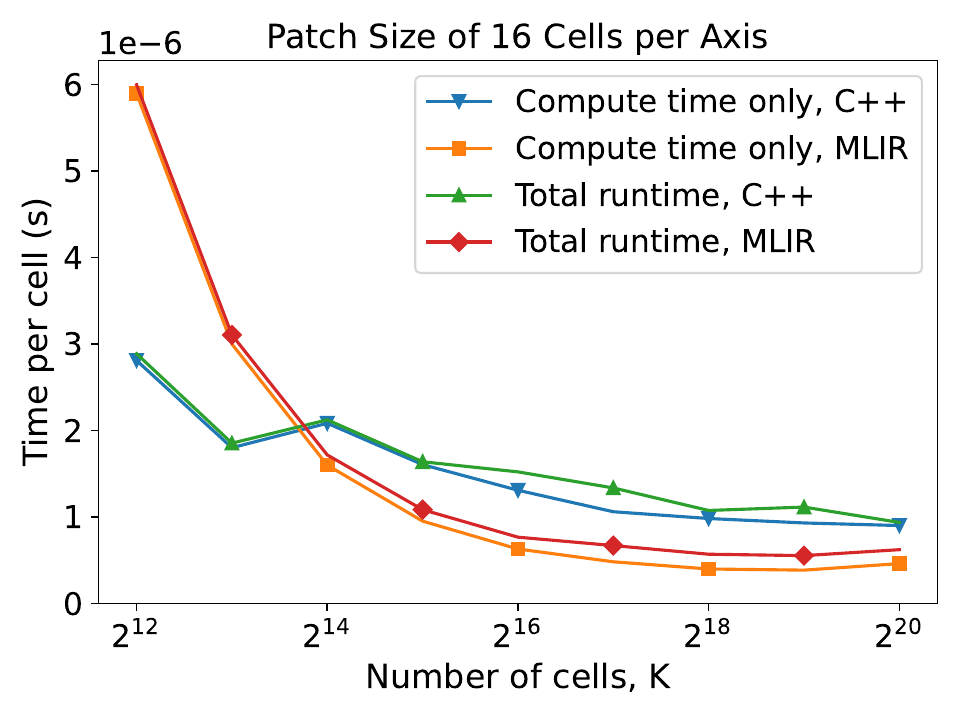}
    \caption{Runtime per cell of CCZ4 with an FD4 solver}
    \label{fig:appendix:plots:ccz4:fd4:gpu_parallel:h200:normalised}
\end{figure}

\begin{figure}[!htb]
    \centering
    \includegraphics[width=0.3\linewidth]{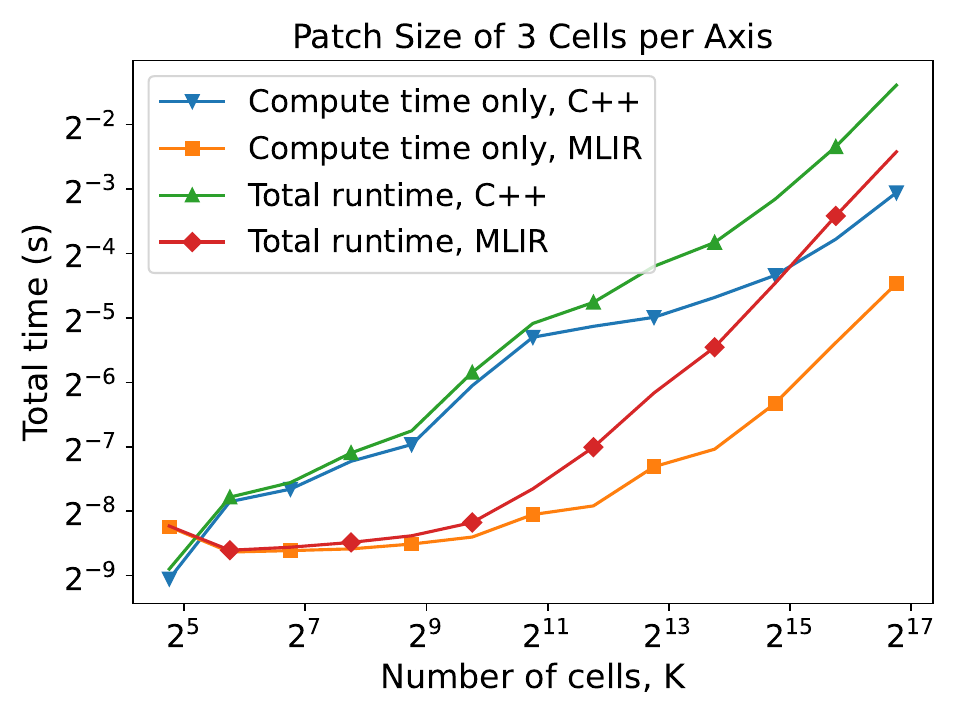}
    \includegraphics[width=0.3\linewidth]{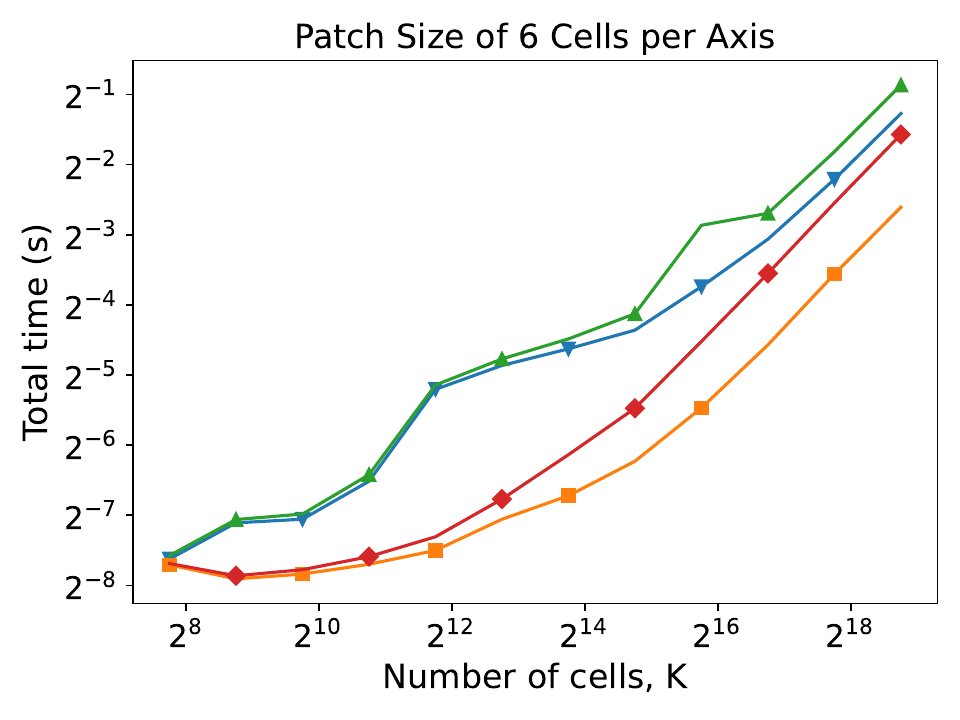}
    \includegraphics[width=0.3\linewidth]{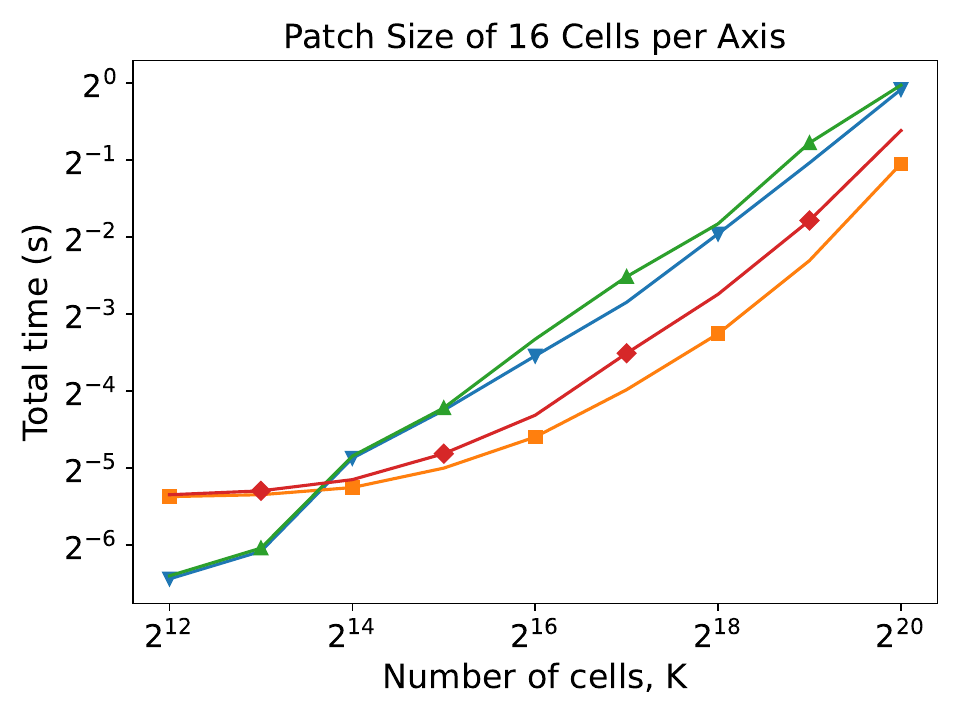}
    \caption{Runtime over all cells of CCZ4 with an FD4 solver}
    \label{fig:appendix:plots:ccz4:fd4:gpu_parallel:h200}
\end{figure}

Here we present plots of our GPU performance for the CCZ4 equations.
Figures \ref{fig:appendix:plots:ccz4:rusanov:gpu_parallel:h200:normalised} and \ref{fig:appendix:plots:ccz4:fd4:gpu_parallel:h200:normalised} show the differences between the compute time and total runtime per cell for the FV and FD4 solvers respectively.
Figures \ref{fig:appendix:plots:ccz4:rusanov:gpu_parallel:h200} and \ref{fig:appendix:plots:ccz4:fd4:gpu_parallel:h200} show the same information presented as the sum over all cells.

\FloatBarrier

\end{document}